\documentclass[10pt,journal]{IEEEtran}

\usepackage{cite}
\usepackage{graphicx}
\usepackage[cmex10]{amsmath}
\usepackage{amsfonts}
\usepackage{amssymb}
\usepackage{mathrsfs}
\usepackage{algorithmic}
\usepackage{algorithm}
\usepackage{array}
\usepackage{mdwmath}
\usepackage{mdwtab}
\usepackage[caption=false,font=footnotesize]{subfig}
\usepackage{fixltx2e}
\usepackage{stfloats}
\usepackage{footnote}
\usepackage{tabularx}
\usepackage{multirow}
\usepackage{color}
\usepackage{textcomp}

\newcounter{tempeqncnt}

\newtheorem{definition}{Definition}
\newtheorem{proposition}{Proposition}

\newtheorem{lemma}{Lemma}

\newtheorem{remark}{Remark}

\begin{document}

\title{Technology Choices and Pricing Policies \\ in Public and Private
 Wireless Networks}

\author{Yuanzhang~Xiao, William~R.~Zame, and~Mihaela~van~der~Schaar,~\IEEEmembership{Fellow,~IEEE}%
\thanks{Y. Xiao and M. van der Schaar are with the Department of Electrical Engineering, University of California, Los Angeles, CA 90095 USA (e-mail: \{yxiao,mihaela\}@ee.ucla.edu).}%
\thanks{W. R. Zame is with the Department of Economics, University of California, Los Angeles, CA 90095 USA (e-mail: zame@econ.ucla.edu).}}

\maketitle

\begin{abstract}
This paper studies the provision of a wireless network by a monopolistic provider who may be either benevolent (seeking to maximize social welfare)
or selfish (seeking to maximize provider profit). The paper addresses questions that do not seem to have been studied before in the engineering
literature on wireless networks: Under what circumstances is it feasible for a provider, either benevolent or selfish, to operate a network in such a
way as to cover costs? How is the optimal behavior of a benevolent provider different from the optimal behavior of a selfish provider?  And, most
importantly, how does the medium access control (MAC) technology influence the answers to these questions? To address these questions, we build a
general model, and provide analysis and simulations for simplified but typical scenarios; the focus in these scenarios is on the contrast between the
outcomes obtained under carrier-sensing multiple access (CSMA) and outcomes obtained under time-division multiple access (TDMA). Simulation results
demonstrate that differences in MAC technology can have a significant effect on social welfare, on provider profit, and even on the (financial)
feasibility of a wireless network.
\end{abstract}

%\begin{IEEEkeywords}

%\end{IEEEkeywords}

\section{Introduction} \label{sec:intro}
There has been much recent debate about the deployment of wireless networks that would allow Internet access in public areas. Central to this debate
is the tradeoff between costs and benefits. Surprisingly, this debate seems to have ignored that the costs and benefits of such wireless networks
depend crucially on the technology that is or could be employed. The purpose of this paper is to provide a framework for exploring the influence of
technology on the costs and benefits of wireless networks and to demonstrate in a simple scenario that the feasibility and profitability of such a
network may depend on the technology chosen.

Although we are most interested in the analysis of {\em public} wireless networks, we construct a framework general enough to allow for the analysis
of {\em private} wireless networks as well. Here, we identify a network as being public if the operator is benevolent, and seeks to maximize social
welfare; we identify a network as being private if the operator is selfish, and seeks to maximize profit. We show that the analysis of both public
and private wireless networks depends crucially on the technology layer, the application layer, and the economic layer, and most crucially of all, on
the interactions between these layers. Indeed, even a proper {\em description} of the environment depends on the interaction between these layers.

To see why the analysis depends crucially on the interactions between the various layers, consider a simple but representative scenario. There are
two classes of (potential) users: {\em email users}, who are insensitive to throughput and delay, and {\em video users}, who are sensitive to both
throughput and delay. In managing the network, the service provider can offer a pricing policy, but the service provider's range of choices depends
on the technology -- in particular, on the medium access control (MAC) protocol -- employed. If time-division multiple access (TDMA) is employed, the
service provider will be able to guarantee the data rate and delay experienced by the users, and thus charge the users according to the guaranteed
data rates. If carrier-sensing multiple access (CSMA) is employed, the service provider will be unable to guarantee the data rate or delay. Absent
such performance guarantees, users may be unwilling to pay for the data rates. As we will show, there are large regions within the range of plausible
parameters in which employing TDMA rather than CSMA makes possible large improvements in social welfare. Indeed, there are regions in which employing
TDMA would be consistent with operating a self-financing network while employing CSMA would not be.

\subsection{Related Work}
Two substantial bodies of work in the engineering literature ask about optimal behavior of the provider of a wireless network. The first considers a
benevolent provider whose objective is to maximize social welfare\cite{PalomarChiang06}--\cite{JohariTsitsiklis09}; the second considers a selfish
provider whose objective is to maximize profit\cite{BasarSrikant02}--\cite{RenParkvanderSchaar_ToN}. While the works in
\cite{PalomarChiang06}--\cite{LiHuang} study the optimal pricing policies and resource allocation strategies of the provider, more recent works
\cite{ManshaeiHubaux}--\cite{RenParkvanderSchaar_ToN} also consider the technology selection problem faced by the provider. Our work is unique in
that we focus on the \emph{comparison} between the optimal behavior of the benevolent provider and the selfish provider in terms of their pricing
policies, and perhaps more importantly, the impact of the technology on the optimal behavior of the benevolent and selfish providers.\footnote{The
interplay of technology and pricing policies is discussed by Lehr \emph{et al.} \cite{SirbuLehrGillett06}, but their paper provides no mathematical
model or quantitative analysis.}

Apart from the focus of the paper, our work differs from existing works in three key elements of the system model. First, we model prices as real
prices actually paid by users and collected by the service providers. However, the prices in some works \cite{PalomarChiang06}--\cite{ShenBasar07}
are not real prices actually paid by the users; rather, they are control signals used for the purpose of controlling the network congestion. Palomar
and Chiang \cite{PalomarChiang06} and Kelly \emph{et al.} \cite{Kelly97}\cite{Kelly98} consider a network with one service provider serving multiple
users and propose charging in proportion to the flow rates of the users in order to maximize social utility. Johari and Tsitsiklis
\cite{JohariTsitsiklis04}\cite{JohariTsitsiklis09} focus on the efficiency loss under this pricing scheme and a variant with price differentiation.
Gibbens and Kelly \cite{GibbensKelly99} propose a packet-based pricing policy for more effective flow control. Under the same scenario, Basar
\emph{et al.} \cite{BasarSrikant02}\cite{ShenBasar04}\cite{ShenBasar07} propose linear and nonlinear differentiated pricing schemes to control the
network usage and maximize the provider's revenue.

Second, we model the users as non-atomic strategic players, who decide whether to enter the network or which pricing plan to choose based on their
utility functions. The price influences the users' decisions, which in turn impact the aggregate arrival rate, i.e. the user demand. However, some
works \cite{PaschalidisTsitsiklis00}--\cite{RenParkvanderSchaar_ToN} use a continuum user model with atomic users that is abstracted by a single
parameter, the user demand, which is simply determined as a function of the price (and maybe the congestion). Paschalidis and Tsitsiklis
\cite{PaschalidisTsitsiklis00} studies a dynamic network with users arriving and leaving the network and derives the optimal pricing strategy and its
static approximation. Similarly, Starobinski \emph{et al.} \cite{MutluAlanyaliStarobinski09}\cite{AlDaoudAlanyaliStarobinski10} studies optimal
pricing strategies in dynamic spectrum access networks.

Third, since we focus on the influence of MAC protocols in a Wireless LAN on the optimal behavior of the providers, we model the technology layer
closely as the MAC layer in wireless networks. We derive analytical expressions for the data rates achieved by MAC protocols in our model, in order
to determine the utility of the users and their payment based on the pricing policies. In addition, the congestion experienced by the users is more
accurately modeled as experienced in a wireless network using the considered MAC protocols. On the contrary, most works in
\cite{PalomarChiang06}--\cite{RenParkvanderSchaar_ToN} consider resource allocation in higher layers (e.g. flow-level resource allocation
\cite{PalomarChiang06}--\cite{ShenBasar07}\cite{LiHuang}) or model the technology as a function that determines the congestion based on the demand
\cite{ManshaeiHubaux}--\cite{RenParkvanderSchaar_ToN}.

Other papers use much different models and have a much different focus. For example, Friedman and Parkes \cite{FriedmanParkes02} study the existence
of implementable mechanisms for the users to truthfully announce their arrivals in WiFi networks. Musacchio and Walrand \cite{MusacchioWalrand06}
model WiFi pricing as a dynamic game involving one access point and one user, and study the Nash equilibrium (NE) of this game. We do not discuss all
of them due to space limit.

The remainder of this paper is organized as follows. In Section II, we introduce the system model for the three-layer network. In Section III, we
formulate the design problem for the benevolent and selfish providers and the decision process of the users as a two-stage game (with the provider
acting in the first stage and the users acting in the second stage). In Section IV, we focus our analysis on a typical scenario to gain insights into
this problem, and provide simulation results in this typical scenario. Finally, Section V states our conclusions.

\section{System Model} \label{sec:model}
We consider a public wireless network, created by a service provider to enable Internet connections to potential users in public areas such as parks,
libraries, and coffee shops. We focus on a wireless local area network (LAN) with a single access point (AP). The model also applies to wireless LANs
with multiple APs to cover a larger area, as long as the APs are operated by the same service provider and share the same spectrum. Keeping in mind
that a wireless LAN will typically serve a relatively small number of potential users who may come and go at any moment in time, we build a dynamic
continuous-time framework in which a finite number of potential users arrive and depart randomly.

In our framework, the system consists of three layers, namely the {\em technology layer}, the {\em application layer}, and the {\em economic layer}.
The technology layer includes the MAC protocol and the admission control policy chosen by the service provider; the application layer includes the
users' utility functions, arrival rates, and service times; and the economic layer includes the pricing plan offered by the service provider and
chosen by the users. The usual way to describe a system model is to describe separately and in turn each of the layers. However, in our settings, it
is not possible to describe these layers separately because they are interconnected. Instead, we describe the system by the specifications for the
service provider and users. In this way, we can better illustrate the interactions among the components in the system and the behaviors of the
service provider and users.

Before we begin with the description of the service provider, we first introduce the basic concept of the user type. The \emph{users} are categorized
into $K$ types according to their utility functions and arrival and departure processes. There are $N_k$ identical users of type $k$.

\subsection{The Service Provider}
The service provider must choose a MAC protocol, a pricing policy, and an admission control policy. (Here we view the MAC protocol as a design choice
of the provider. However, our analysis shows that, even if the MAC protocol is \emph{not} a design choice, the MAC protocol that is used has an
important impact on the remainder of the design and performance and in particular on the feasibility of providing service.) However, before
describing these three design parameters, we must note an important {\em caveat}. A MAC protocol describes which packets of current users will have
access to which resources in which way; an admission control policy describes which users will be admitted to the system. It would seem that a MAC
protocol should allocate resources to packets of current users (or type of users), as a function of the current number of users of each type in the
system, and that an admission policy should specify whether or not an incoming user of a particular type should be admitted to the system, as a
function of the current number of users of each type in the system. However, the service provider can only use policies that depend on observable
characteristics and actions of the users and {\em the type of a user cannot be observed by the service provider}.\footnote{Leaving aside the point
that users might lie about their types in order to obtain more favorable treatment, it might simply be the case that there are more types of users
than there are pricing plans, so that different types of users necessarily choose the same pricing plan.} In our framework, the relevant observable
actions of the users are their choices of pricing plans, so the policies of the service provider should be specified as functions of the choices of
pricing plans.

\subsubsection{The Medium Access Control Protocol}
The MAC protocol chosen determines the ways in which users may share the channel resources. In principle, the service provider might be able to
choose among many MAC protocols. CSMA and TDMA are the canonical MAC protocols. CSMA is representative of the protocols without a central controller,
where the packets contend to get access to the medium. The widely-used IEEE 802.11 standards use CSMA as the basic MAC protocol\cite{80211b99}. TDMA
is representative of the protocols with a central controller, where the packets access the medium in non-overlapping periods of time. The IEEE
802.11e standard enables contention-free access control in the Hybrid Control Function (HCF), which can be considered as a generalized TDMA
protocol\cite{80211e04}. The key difference between CSMA and TDMA is that TDMA enables the provider to offer quality of service (QoS) guarantee,
while CSMA does not. Specifically, the users can guarantee to achieve a certain data rate in TDMA, while their data rates may vary greatly because of
the probabilistic channel access in CSMA. Hence, it is impossible to charge the users based on their guaranteed data rates in CSMA. On the contrary,
a provider using TDMA can charge users based on their guaranteed data rates. We write $\theta$ for a particular protocol.

\subsubsection{Pricing plans, Pricing Policies, and Pricing States}
A {\em pricing plan} is a schedule of charges to (potential) users. For simplicity, we assume that charges consist only of a subscription fee (paid
once per billing period) $p_s$ and a charge $q$ per unit for the guaranteed data rate.\footnote{There would be no difficulty in allowing connection
fees, fees that depend on minutes of usage, fees that depend on time of day, etc. We focus here on a simpler model to make our essential points.} Of
course, the subscription fee or the charge for the guaranteed data rate might be 0. Thus a pricing plan is
$$
{\bf p} = (p_s, q),~p_s\ge0,~q\ge0.
$$
Note that the charge for the guaranteed data rate is applied to the data rate allocated to a user, instead of the user's actual amount of data usage,
in a billing period. This is reasonable because the user should pay for the bandwidth exclusively allocated to its data, even though it does not use
the bandwidth all the time. In CSMA, the bandwidth is not exclusively allocated to a certain user. Instead, the users access the channel
opportunistically and may get extremely high or low effective bandwidth. We take account of this by distinguishing the set ${\mathcal P}_\theta$ of
pricing plans that can be employed given the MAC protocol $\theta$.

To allow for the possibility that some users choose not to belong to the network at all, we will require that the service provider always offer a
dummy plan $\phi$. A user choosing $\phi$ does not subscribe to the network.

A {\em pricing policy} is a vector of pricing plans; for simplicity, we assume here that each pricing policy is a vector of exactly $L+1$ pricing
plans: ${\bf P}_\theta = ({\bf p}^0, {\bf p}^1, \ldots, {\bf p}^L)$; by convention we assume that ${\bf p}^0 = \phi$. Write
$$
{\mathcal P}_\theta^{L+1} = \{\phi\} \times \underbrace{{\mathcal P}_\theta \times \ldots \times {\mathcal P}_\theta}_{L~{\rm times}}
$$
for the set of all possible pricing policies given the MAC control protocol $\theta$.

Given a pricing policy ${\bf P} = ({\bf p}^0, {\bf p}^1,\ldots, {\bf p}^L)$, each user type $k$ chooses a pricing plan from ${\bf P}$ by randomizing
over all the choices according to a probability distribution. We define the {\em pricing state} to be the vector ${\bf v} = (v^0, v^1, \ldots, v^L)$,
where $v^\ell$ is the number of users who are currently online and choose the pricing plan ${\bf p}^\ell$. We write ${\mathcal V}$ for the set of
pricing states.

\subsubsection{The Admission Control Policy}
The admission control policy determines whether a user of a given type should be admitted into the system given the current number of users of each
type in the system. However, as we have noted, the service provider cannot observe the types of the current users or of the incoming user, but only
the pricing plans they choose, so the admission control policy must depend only on the pricing plans chosen by current users -- that is, on what we
have called the pricing state -- and on the pricing plan chosen by the incoming user. Hence, an admission control policy is a function
$$
\alpha: {\mathcal V} \times \{0, 1, \ldots, L\} \to \{ 0,1 \},
$$
where we interpret $1$ to mean 'admit' and $0$ to mean 'do not admit'.

\begin{figure}
\centering
\includegraphics[width =3.5in]{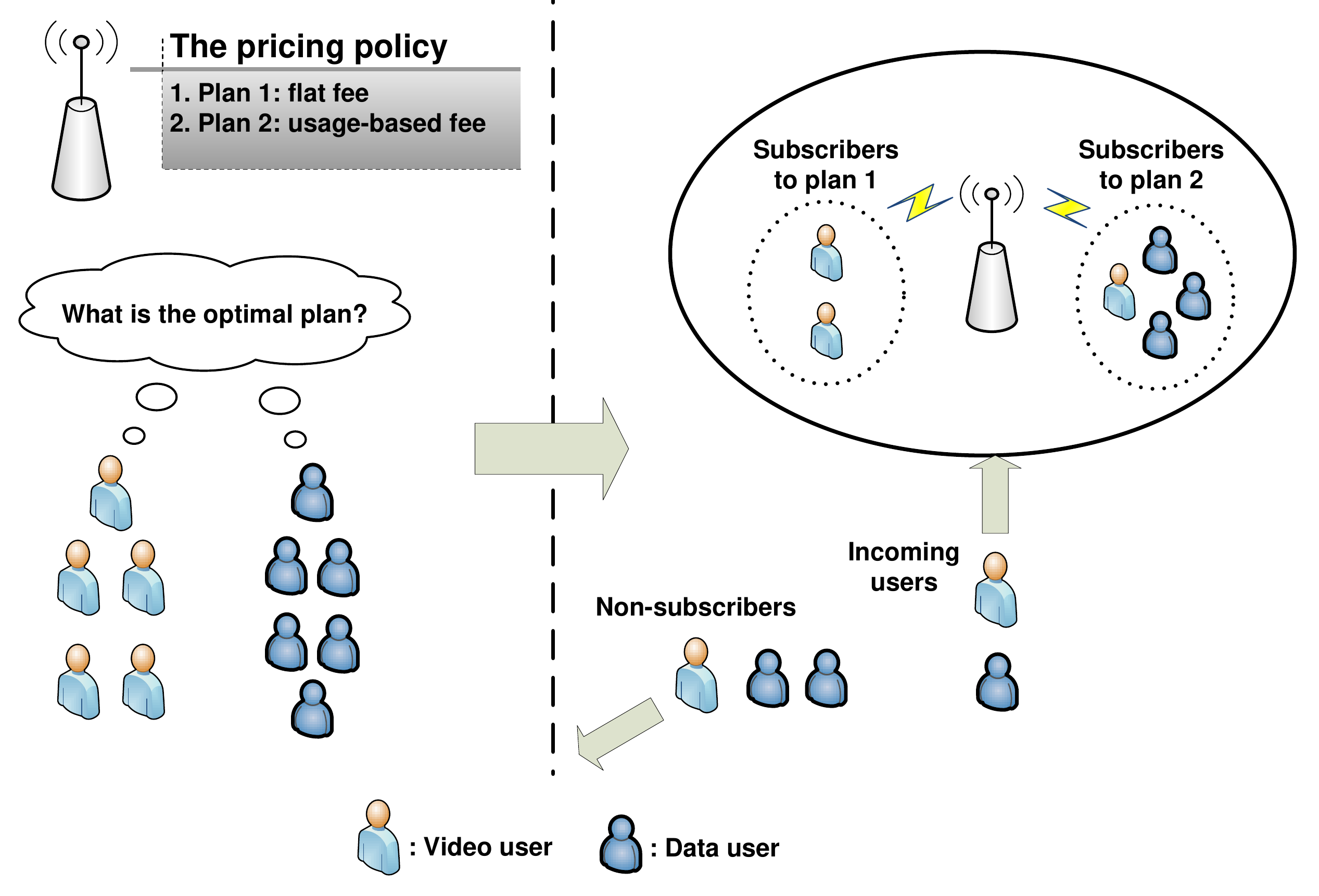}
\caption{Illustration of the system: first, the service provider announces the pricing policy and the users of each type choose the optimal
probability distribution over the pricing plans; then, each user randomizes according to the probability distribution, and subscribes to a particular
plan (or leaves the network).} \label{SystemModel}
\end{figure}

The simplest admission control policy admits every potential user: $\alpha(v,\ell) \equiv 1$ for all $\ell$. A slightly more complicated admission
control policy sets  bounds $\hat{N}_\ell$ for each $\ell = 1, \ldots, L$ and admits a potential user who chooses the pricing plan $\ell$ if and only
if doing so does not cause the total number of current users who have chosen plan $\ell$ to exceed the the upper bound $\hat{N}_\ell$:
$\alpha(v,{\ell}) = 1$ if and only if $v_\ell \leq \hat{N}_\ell - 1$. The upper bound $\hat{N}_\ell$ can be determined according to the total
bandwidth available and the bandwidths allocated to the users with different pricing plans.

\subsection{Users}

The users are characterized by their utility functions, arrival processes, and service times. Given user characteristics and the technology and the
pricing policy adopted by the service provider, each user determines a probability distribution on the choices of pricing plans that maximizes its
expected utility (which will depend on the choices of all the other users). At the beginning of time, each user chooses a pricing plan randomly
according to the prescribed probability distribution, and sticks to the chosen plan throughout the considered time horizon. Every time a user arrives
at the network, the service provider observes the pricing plan chosen by this user. The service provider will make the admission control and medium
access control according to the current pricing state and the choice of a particular user. See Fig.~\ref{SystemModel} for an illustration of an
example system with two types of users.

\subsubsection{Choices of Pricing Plans}
Users choose pricing plans to maximize their expected utility, given the menu of pricing plans, the MAC protocol and the admission control policy of
the provider, and the choices of other users. We allow for the possibility that users {\em randomize} to choose a pricing plan at the beginning of
time. We write $\pi_{k,\ell}$ for the probability that a user of type $k$ chooses plan $\ell$; in particular, $\pi_{k,0}$ is the probability of
choosing the dummy plan $0$. For each $k$, we have $\sum_{\ell = 0}^L \pi_{k,\ell} = 1$. Write $\pi_k = \left[\pi_{k,0}, \ldots, \pi_{k,L}\right]$
for the (random) action of type-$k$ users, $\pi = (\pi_1, \ldots, \pi_K)$ for the action profile of all the users, and $\pi_{-k}$ for the action
profile of users of types other than $k$. Allowing for randomization guarantees that equilibrium exists. We may interpret randomization literally:
users who are indifferent over various plans break their indifference in a random way. Alternatively, we may interpret randomization simply as
uncertainty in the minds of the provider and other users. If the number of users is large, we can also interpret the probability distribution over
pricing plans as the distributions of plans among the population \cite{TembineAltman09}.

The randomization is realized at the beginning of time. Represent the result of the randomization by a set of vectors
$$
{\bf n} = ({\bf n}_1,\ldots,{\bf n}_K)=\left([n_{1,0},\ldots,n_{1,L}],\ldots,[n_{K,0},\ldots,n_{K,L}]\right)
$$
with $n_{k,\ell}$ being the number of type-$k$ users choosing plan $\ell$.

\subsubsection{System State}
The \emph{system state}, or the \emph{true state}, is defined as the number of online users of each type choosing each pricing plan. Specifically,
the system state $\mathbf{x}$ is a $K\times (L+1)$ matrix, with $x_{k,l}$ as the element at the $k$th row and $(l+1)$th column, representing the
number of type-$k$ users who choose plan $l$ and are currently in the network. The system state $\mathbf{x}$ is in $\mathcal{X}_\alpha({\bf n})$, the
set of admissible system states under the admission control policy $\alpha$ and the result of the randomization ${\bf n}$.

The system state depends on arrivals and departures of users and on the admission control policy of the service provider, and thus is random. We
write ${\bf X}(t)$ for the stochastic process of system state evolution. We assume that the type of a user is a private characteristic, known to the
user but generally unobservable by other users and the service provider. As a result, the system state cannot be observed by anyone in the system.

\subsubsection{Arrival Process and Service Time}
For simplicity, we assume the arrival process and service time are exogenously given but not choice variables.\footnote{Here, the arrival process
characterizes the arrival of users, but not the arrival of users' packets. Similarly, the service time is the duration of users staying in the
system.} We use a continuous-time model (reflecting the fact that users might arrive/depart at any moment); as in \cite{RossTsang89}, we assume that
the users arrive independently and the arrival process of type-$k$ users choosing plan $\ell$ is Poisson with arrival rate
$$\lambda_{k,\ell}(t)=\lambda_k\cdot(n_{k,\ell}-x_{k,\ell}(t)),
$$
where $\lambda_k$ is the individual arrival rate of a type-$k$ user. Note that the aggregate arrival rate $\lambda_{k,\ell}(t)$ is proportional to
the number of users currently outside the network $(n_{k,\ell}-x_{k,\ell}(t))$. We assume that the service time of one type-$k$ user is exponentially
distributed with mean $1/\mu_k$, and that different users leave the network independently. Hence, the aggregate departure rate of type-$k$ users
choosing plan $\ell$ is
$$\mu_{k,\ell}(t)=\mu_k\cdot x_{k,\ell}(t).
$$
Note that our results, based on the analysis of the steady state, hold true for other
probability distributions of the arrival and departure processes.

\subsubsection{Billing Period}
We fix a {\em billing period} of length $\Delta T$, which is typically one month. Subscription fees are charged at the beginning of each billing
period; other fees are charged at the end of each billing period. This is consistent with the usual billing methods: people pay a subscription fee
prospectively and other charges retrospectively. For convenience, we assume that neither the provider nor the users discount utility and cost over
the billing period.

\subsubsection{Expected Utility}
The service provider and the users evaluate the social welfare and their satisfaction, respectively, by the \emph{expected utility}, defined as the
expectation of the total utility over a billing period when the stochastic process of the system state ${\bf X}(t)$ reaches the steady state. Each
user's total utility consists of two components: utility of use and disutility of cost. To keep the model simple, we assume that total utility is
simply the sum of utility of use and disutility of cost and is linear in cost with marginal utility of cost equal to $1$\cite{MWGMicroeconomics95}:
\begin{eqnarray}
\mbox{ total utility } = \mbox{ utility of use } - \mbox{ cost }.
\end{eqnarray}

We denote the expected utility of use of a type-$k$ user by $U_k(\theta,\alpha,\pi)$, if the MAC protocol is $\theta$, the admission control policy
is $\alpha$, and the joint probability distribution over pricing plans is $\pi$. We can calculate the expected utility of use
$U_k(\theta,\alpha,\pi)$ as follows
\begin{eqnarray}\label{ExpectedUtilityOfUse}
U_k(\theta,\alpha,\pi) = \sum_{\ell=1}^L \pi_{k,\ell}\cdot\sum_{{\bf n}:n_{k,\ell}\geq1} \Pr({\bf n}|k,\ell)\cdot V_k^\ell(\theta,\alpha,{\bf n}),
\end{eqnarray}
where $\Pr({\bf n}|k,\ell)$ is the conditional probability that the randomization results in ${\bf n}$ given that one type-$k$ user chooses plan
$\ell$ after randomization, and $V_k^\ell(\theta,\alpha,{\bf n})$ is the steady-state utility of use of a type-$k$ user, if the MAC protocol is
$\theta$, the admission control policy is $\alpha$, and the realization of the randomization is ${\bf n}$. %Basic combinatorics knowledge gives us the
%probability $\Pr({\bf n}|k,\ell)$ as
%\begin{eqnarray}\label{ProbabilityOfRandomizationResult}
%\Pr({\bf n}|k,\ell)&=& \binom {N_k-1} {n_{k,\ell}-1}\cdot\pi_{k,\ell}^{n_{k,\ell}-1}\cdot\prod_{\ell^\prime=0,\ell^\prime\neq\ell}^L \left({\binom {N_k-\sum_{m=0,m\neq\ell}^{\ell^\prime-1}n_{k,m}-n_{k,\ell}} {n_{k,\ell^\prime}}}\cdot\pi_{k,\ell^\prime}^{n_{k,\ell^\prime}}\right)\nonumber \\
%&\cdot&\prod_{k^\prime=1,k^\prime\neq k}^K\prod_{\ell^\prime=0}^L\left(\binom {N_{k^\prime}-\sum_{m=0}^{\ell^\prime-1}n_{k^\prime,m}}
%{n_{k^\prime,\ell^\prime}}\cdot\pi_{k^\prime,\ell^\prime}^{n_{k^\prime,\ell^\prime}}\right),
%\end{eqnarray}
%where $\binom {n} {k}$ is the number of $k$-element subsets of an $n$-element set.

\begin{figure*}[!t]
\normalsize \setcounter{tempeqncnt}{\value{equation}} \setcounter{equation}{\value{equation}}
\begin{align}
\label{SteadyStateUtilityOfUse} V_k^\ell(\theta,\alpha,{\bf n}) = & \lim_{m\to\infty} \int_{m \Delta T}^{(m+1) \Delta T} \!\!\!\! \sum_{{\bf
x}\in\mathcal{X}_\alpha({\bf n})} \Pr({\bf X}(t)=\mathbf{x})\cdot
\frac{x_{k,\ell}}{n_{k,\ell}} \cdot u_k(\tau^\theta_{k,\ell}(\mathbf{x}),\delta^\theta_{k,\ell}(\mathbf{x}))\cdot dt \nonumber \\
=&~\Delta T \cdot \lim_{t\to\infty} \sum_{\mathbf{x}\in\mathcal{X}_\alpha({\bf n})} \Pr({\bf X}(t)=\mathbf{x}) \cdot
\frac{x_{k,\ell}}{n_{k,\ell}}\cdot
u_k(\tau^\theta_{k,\ell}(\mathbf{x}),\delta^\theta_{k,\ell}(\mathbf{x})) \\
=&~\Delta T \cdot \sum_{\mathbf{x}\in\mathcal{X}_\alpha({\bf n})} \Pr({\bf X}(\infty)=\mathbf{x}) \cdot \frac{x_{k,\ell}}{n_{k,\ell}}\cdot
u_k(\tau^\theta_{k,\ell}(\mathbf{x}),\delta^\theta_{k,\ell}(\mathbf{x})), \nonumber
\end{align}
\hrulefill \vspace*{4pt} \setcounter{equation}{\value{tempeqncnt}+1}
\end{figure*}

The steady-state utility of use $V_k^\ell(\theta,\alpha,{\bf n})$ given ${\bf n}$ is shown in \eqref{SteadyStateUtilityOfUse} at the top of the page,
%\begin{eqnarray}
%\label{SteadyStateUtilityOfUse} V_k^\ell(\theta,\alpha,{\bf n}) & = & \lim_{m\to\infty} \int_{m \Delta T}^{(m+1) \Delta T} \sum_{{\bf
%x}\in\mathcal{X}_\alpha({\bf n})} \Pr({\bf X}(t)=\mathbf{x})\cdot
%\frac{x_{k,\ell}}{n_{k,\ell}} \cdot u_k(\tau^\theta_{k,\ell}(\mathbf{x}),\delta^\theta_{k,\ell}(\mathbf{x}))\cdot dt \nonumber \\
%&=&\Delta T \cdot \lim_{t\to\infty} \sum_{\mathbf{x}\in\mathcal{X}_\alpha({\bf n})} \Pr({\bf X}(t)=\mathbf{x}) \cdot
%\frac{x_{k,\ell}}{n_{k,\ell}}\cdot
%u_k(\tau^\theta_{k,\ell}(\mathbf{x}),\delta^\theta_{k,\ell}(\mathbf{x})) \\
%&=&\Delta T \cdot \sum_{\mathbf{x}\in\mathcal{X}_\alpha({\bf n})} \Pr({\bf X}(\infty)=\mathbf{x}) \cdot \frac{x_{k,\ell}}{n_{k,\ell}}\cdot
%u_k(\tau^\theta_{k,\ell}(\mathbf{x}),\delta^\theta_{k,\ell}(\mathbf{x})), \nonumber
%\end{eqnarray}
where $\Pr({\bf X}(t)=\mathbf{x})$ is the probability that the current system state ${\bf X}(t)$ is $\mathbf{x}$, $u_k$ is the instantaneous utility
of use of a type-$k$ user, and $\tau^\theta_{k,\ell}(\mathbf{x})$ and $\delta^\theta_{k,\ell}(\mathbf{x})$ are the throughput and delay of a type-$k$
user choosing plan $\ell$, respectively, if the user is online, the MAC protocol is $\theta$, and the system state is $\mathbf{x}$. Since there are a
finite number of reversible system states, the steady state of the process ${\bf X}(t)$ and thus the limit in (\ref{SteadyStateUtilityOfUse}) always
exist. The system state at the steady state is a random variable $\mathbf{X}(\infty)$, and its distribution $\Pr({\bf X}(\infty)=\mathbf{x})$ can be
calculated analytically as
\begin{eqnarray}\label{ProbabilityOfSystemState}
\Pr({\bf X}(\infty)=\mathbf{x}) = \Pr({\bf X}(\infty)=\mathbf{0}) \cdot \!\!\!\prod_{k=1}^K \prod_{\ell=0}^L \prod_{j=1}^{x_{k,\ell}} \frac{\lambda_k
\cdot (n_{k,\ell}-j)}{\mu_k\cdot j}, \nonumber
\end{eqnarray}
where $\Pr({\bf X}(\infty)=\mathbf{0})$ is the probability of system state $\mathbf{0}\in\mathbb{R}^{K\times(L+1)}$, in which no user is in the
system, and is determined by $\sum_{\mathbf{x}\in\mathcal{X}_\alpha({\bf n})} \Pr({\bf X}(\infty)=\mathbf{x}) = 1$.

%We assume that $u_k$'s are bounded and continuous; they need not be concave in the throughput or delay. We normalize so that users who are not online
%experience instantaneous utility of use $0$. Similarly, users who are online but experience $0$ throughput or infinite delay also experience
%instantaneous utility $0$: $u_k(0,*) = u_k(*, \infty) = 0$.

We denote the expected cost of a type-$k$ user by $C_k(\theta,\alpha,{\bf P},\pi)$, if the MAC protocol is $\theta$, the admission control policy is
$\alpha$, the pricing policy is ${\bf P}$, and the joint probability distribution over pricing plans is $\pi$. The expected cost
$C_k(\theta,\alpha,{\bf P},\pi)$ can be calculated by
\begin{eqnarray}\label{ExpectedCost}
\sum_{\ell=1}^L \pi_{k,\ell} \left(p_s^\ell + \sum_{{\bf n}:n_{k,\ell}\geq1} \Pr({\bf n}|k,\ell)\cdot q^\ell\cdot B_k^\ell(\theta,\alpha,{\bf
n})\right),
\end{eqnarray}
where $B_k^\ell(\theta,\alpha,{\bf n})$ is the expected amount of data rates guaranteed for a type-$k$ user choosing plan $\ell$ over a billing
period at the steady state, shown as below
\begin{eqnarray}\label{DataRateGivenUserNumber}
&& B_k^\ell(\theta,\alpha,{\bf n}) \nonumber \\
&=& \lim_{m\to\infty}\int_{m \Delta T}^{(m+1) \Delta T} \!\!\!\!\sum_{\mathbf{x}\in\mathcal{X}_\alpha(\mathbf{n})} \!\!\!\!\Pr({\bf
X}(t)=\mathbf{x})\cdot
\frac{x_{k,\ell}}{n_{k,\ell}} \cdot \tau^\theta_{k,\ell}(\mathbf{x})\cdot dt \nonumber \\
&=& \Delta T \cdot \sum_{\mathbf{x}\in\mathcal{X}_\alpha(\mathbf{n})} \mathrm{Pr}(\mathbf{X}(\infty)=\mathbf{x})\cdot \frac{x_{k,\ell}}{n_{k,\ell}}
\cdot \tau^\theta_{k,\ell}({\bf x}).
\end{eqnarray}

According to our definition, the expected utility is the expected utility of use minus the expected cost
$$U_k(\theta,\alpha,\pi)-C_k(\theta,\alpha,{\bf P},\pi).$$

\subsubsection{Users' Decision Process}
Each user determines the randomizing probability that maximizes its own expected utility given the other users' actions. Consider the decision
process of a particular type-$k$ user. For convenience, we write $(\pi;\pi_k^\prime)$ for the action profile in which the considered type-$k$ user
chooses $\pi_k^\prime$, the other type-$k$ users choose $\pi_k$, and the users of types other than $k$ choose $\pi_{-k}$. When having
$(\pi;\pi_k^\prime)$, instead of $\pi$, as a variable, $U_k(\theta,\alpha,(\pi;\pi_k^\prime))$ and $C_k(\theta,\alpha,{\bf P},(\pi;\pi_k^\prime))$
denote the utility of use and cost of the considered type-$k$ user, respectively. $U_k(\theta,\alpha,(\pi;\pi_k^\prime))$ can be calculated by
\begin{eqnarray}\label{ExpectedUtilityOfUseOfParticularUser}
\sum_{\ell=1}^L \pi_{k,\ell}^\prime\cdot\sum_{{\bf n}:n_{k,\ell}\geq1} \Pr({\bf n}|k,\ell)\cdot V_k^\ell(\theta,\alpha,{\bf n}),
\end{eqnarray}
and $C_k(\theta,\alpha,{\bf P},(\pi;\pi_k^\prime))$ can be calculated by
\begin{eqnarray}\label{ExpectedCostOfParticularUser}
\sum_{\ell=1}^L \pi_{k,\ell}^\prime \left(p_s^\ell + \sum_{{\bf n}:n_{k,\ell}\geq1} \Pr({\bf n}|k,\ell)\cdot q^\ell\cdot B_k^\ell(\theta,\alpha,{\bf
n})\right).
\end{eqnarray}
Note that when $\pi_k^\prime=\pi_k$, we have $U_k(\theta,\alpha,(\pi;\pi_k^\prime))=U_k(\theta,\alpha,\pi)$ and $C_k(\theta,\alpha,{\bf
P},(\pi;\pi_k^\prime))=C_k(\theta,\alpha,{\bf P},\pi)$. Since each user maximizes their own expected utility given the others' decisions, we model
the user interaction as the \emph{plan selection game} defined as
$$
\mathcal{G}_{\bf P}=\left\{\{1,\ldots,K\}, \{\pi_k\}_{k=1}^K, \{U_k-C_k\}_{k=1}^K\right\}.
$$
Here we put ${\bf P}$ in the subscript of $\mathcal{G}$ to emphasize that the plan selection game depends on the pricing policy of the provider.

The outcome of the users' decision process is naturally the Nash equilibrium of the plan selection game defined as follows.
\begin{definition}\label{def:NE}
$\pi$ is a (symmetric) Nash equilibrium of the game $\mathcal{G}_{\bf P}$ if for all $k$,
\begin{eqnarray}\label{UserDecisionProcess}
\pi_k \in \arg \max_{\pi_k^\prime} \left\{U_k(\theta,\alpha,(\pi;\pi_k^\prime))-C_k(\theta,\alpha,{\bf P},(\pi;\pi_k^\prime))\right\},
\end{eqnarray}
\end{definition}
Since we use $\pi_k$ for the action of all the type-$k$ users, the NE defined above is a symmetric NE.

\section{Problem Formulation}
In this section, we formulate the interaction between the service provider and the users as a Stackelberg game. The service provider chooses a MAC
protocol, a pricing policy, and an admission control policy, foreseeing Nash equilibria of the plan selection game played by the users. The design
problem of the service provider is therefore to find a MAC protocol $\theta$, a pricing policy ${\bf P}$, and an admission control policy $\alpha$,
so that at an equilibrium of the plan selection game $\mathcal{G}_{\bf P}$, the social welfare (for the benevolent provider) or the total revenue
(for the selfish provider) is maximized, subject to the constraint that costs be covered.

Note that our notion of solution assumes that the service provider knows the arrival rates, service times, and utility functions of all types of
users (but does not know the type of a particular user), and foresees the behavior of the users. The users in turn must also know the behavior of
other users. Implicitly, therefore, we view the outcome as involving some learning process that is not modeled here. We intend to address this issue
in later work, while focusing on characterizing the system performance at the equilibria in this paper.

Under the above assumptions, we can formulate the design problem of the service provider as follows. For a benevolent service provider aiming at
maximizing the social welfare, its design problem can be written as
\begin{eqnarray}
\!\!\!\!\!\!&\displaystyle\max_{\theta,{\bf P},\alpha}& S(\theta,\alpha,{\bf P},\pi)\triangleq\sum_{k=1}^K \left(U_k(\theta,\alpha,\pi)-C_k(\theta,\alpha,{\bf P},\pi)\right)\cdot N_k \nonumber \\
\!\!\!\!\!\!&s.t.& \pi~\mathrm{is~a~NE~of~the~plan~selection~game}~\mathcal{G}_{\mathbf{P}}, \nonumber \\
\!\!\!\!\!\!&    & \sum_{k=1}^K C_k(\theta,\alpha,{\bf P},\pi)\cdot N_k \geq C_0\cdot 1_{\{\exists k:\pi_{k,0}<1\}},
\end{eqnarray}
where $S(\theta,\alpha,{\bf P},\pi)$ is the social welfare defined as the sum utility of all the users, $C_0$ is the fixed cost for the service
provider during a billing period due to the maintenance of the network, and $1_{\{A\}}$ is the indicator function of the event $A$. The second
constraint is the individual rationality constraint for the service provider, which says that the provider needs a revenue large enough to cover the
cost of running the network. However, if all the users choose the dummy plan, i.e. $\pi_{k,0}=1$ for all $k$, the network does not operate, the
provider has no cost and revenue, and the social welfare will be 0. The solution ${\bf P}^*$ to the above problem provides the users with a set of
pricing plans to choose from. Note that for all the optimization problems in this paper, the optimal solution is $\emptyset$ and the optimal value is
$-\infty$ if the problem is infeasible.

Similarly, for a selfish service provider aiming at maximizing its own revenue, its design problem can be written as
\begin{eqnarray}
&\displaystyle\max_{\theta,{\bf P},\alpha}& R(\theta,\alpha,{\bf P},\pi)\triangleq\sum_{k=1}^K C_k(\theta,\alpha,{\bf P},\pi)\cdot N_k \nonumber \\
&s.t.& \pi~\mathrm{is~a~NE~of~the~plan~selection~game}~\mathcal{G}_{\mathbf{P}}, \nonumber \\
&    & R(\theta,\alpha,{\bf P},\pi) \geq C_0\cdot 1_{\{\exists k:\pi_{k,0}<1\}},
\end{eqnarray}
where $R(\theta,\alpha,{\bf P},\pi)$ is the revenue defined as the total payment of all the users.

Because our focus is the influence of technology on the economic layer and system performance, we will first solve the design problems of the
providers with a fixed MAC protocol and admission control policy, and then compare the optimal pricing policies and the resulting system performance
under different MAC protocols and admission control policies.

Even for a fixed technology, the general design problem for the service provider may not be easy to solve because the provider must, in principle,
foresee the Nash equilibrium behavior of users in the plan selection game following all possible pricing policies and must take into account that
such Nash equilibrium might not be unique.\footnote{In the cases of multiple equilibria, it is not difficult to construct an incentive scheme that
implements a desired action profile in Nash equilibria. Please see Appendix A of Chapter 23 in \cite{MWGMicroeconomics95} for details. For
simplicity, here we assume that the provider can choose the best Nash equilibrium with respect to its objective function.} As we shall see, however,
the design problem is tractable in some settings that provide useful insights.

Before moving to the detailed analysis, we first guarantee the existence of Nash equilibrium in the general settings.
\begin{proposition}
In the plan selection game $\mathcal{G}_{\bf P}$, there always exists a Nash equilibrium as defined in Definition~\ref{def:NE}.
\end{proposition}
\begin{IEEEproof}
The plan selection game $\mathcal{G}_{\bf P}$ is a finite game; Nash shows that such a game has a Nash equilibrium \cite[Theorem~1]{Nash51}.
Moreover, since the users of the same type are symmetric, there exists a symmetric Nash equilibrium in which the users of the same type choose the
same action \cite[Theorem~2]{Nash51}.
\end{IEEEproof}

\section{Detailed Analysis on Two Typical Scenarios}
\begin{figure}
\centering
\includegraphics[width =3.3in]{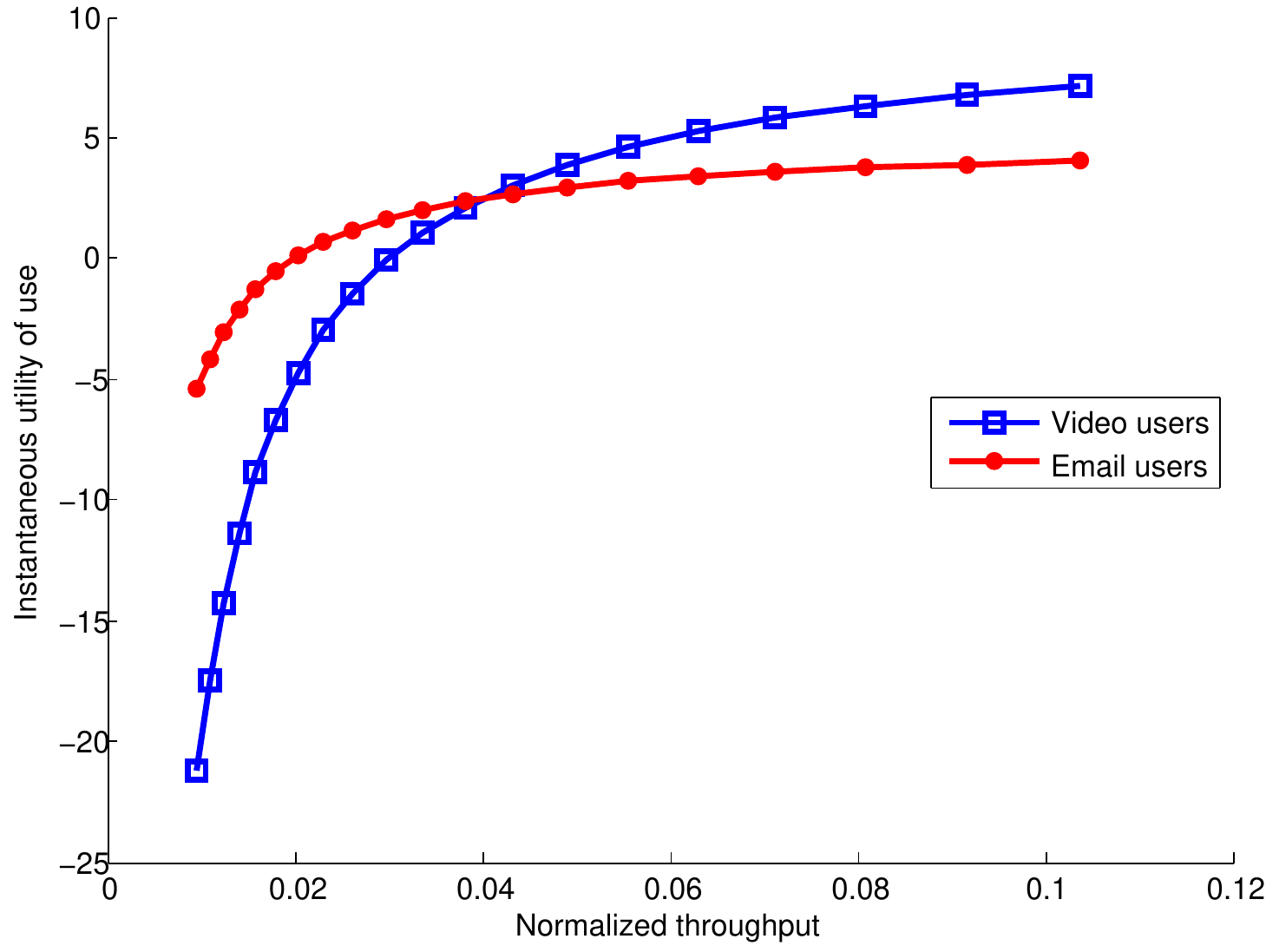}
\caption{Example instantaneous utility of use for the video and email users. For video users, we have $\alpha_1=10,\beta_1=0.3$. For email users, we
have $\alpha_2=5,\beta_2=0.1$. The normalized throughput is calculated in CSMA with the number of users ranging from 20 to 1. The parameters in the
CSMA protocol are chosen such that the throughput is the same as that of a slotted ALOHA protocol with channel access probability $2/17$.}
\label{ExampleUtility}
\end{figure}

In this section, we study two typical scenarios. In both scenarios, there are two types of users: type-1 users are video users with stringent
throughput and delay requirements, while type-2 users are email users, who require low throughput and can tolerate large delay. We assume that the
instantaneous utility of type-$k$ users is a concave function of the form $u_k(\tau) = \alpha_k-\frac{\beta_k}{\tau}$, where $\tau$ is the
throughput. $\alpha_k>0$ represents the highest instantaneous utility a type-$k$ user can get, and $\beta_k$ reflects the rate of increase of utility
with respect to the throughput. See Fig.~\ref{ExampleUtility} for an illustration. Although not essential for the analysis, the following two
assumptions are made. We first assume $\alpha_1>\alpha_2$ to reflect the fact that video users can get higher utility if the throughput is large. We
also assume $\beta_1/\alpha_1>\beta_2/\alpha_2$, because video users need a higher throughput to get positive utility. Note that the delay is not
included in the utility function for simplicity, because higher throughput comes with lower delay in the CSMA and TDMA protocols considered in our
paper. Thus, the user preference can be characterized by a utility function dependent solely on the throughput.

In the first scenario, the service provider uses CSMA, cannot guarantee the data rate of a specific user, and can only offer a pricing plan with a
subscription fee alone. In the second scenario, the service provider uses TDMA, can guarantee the data rate of each user, and can offer a plan with a
subscription fee and a charge proportional to the date rate. We characterize the system performance at the equilibria in both scenarios. Our focus is
on understanding how the equilibrium and network performance are affected by the service provider's objective and the technology adopted.

\subsection{CSMA}
The provider using CSMA offers the dummy pricing plan ${\bf p}^0 = \phi$ and a single non-dummy pricing plan\footnote{There is no need to offer more
than one non-dummy pricing plans, because the users always prefer the non-dummy pricing plan with the lowest subscription fee.} ${\bf p}^1 = (p_s,
0)$. The design problem of the provider can be analyzed using backward induction. In the plan selection game, there are nine types of Nash equilibria
depending on the value of $\pi_{k,1}$: $\pi_{k,1}=0$, $\pi_{k,1}=1$, or $\pi_{k,1}\in(0,1)$. The benevolent (selfish) provider compares the social
welfare (revenue) achievable at all the possible equilibria and adopts the pricing policy that induces the NE with the highest social welfare
(revenue).

To distinguish user behavior in different types of NE, we use the superscript $(t_1,t_2)$ to denote the type of NE, where $t_k=i, o, m~(k=1,2)$
corresponds to the case of $\pi_{k,1}=1$ (type-$k$ users are all `in' the network), the case of $\pi_{k,1}=0$ (they are all `out' of the network),
and the case of $\pi_{k,1}\in(0,1)$ (they are `mixed'), respectively. For example, $\pi^{(i,m)}$ denotes the action profile in which type-$1$ users
are in ($\pi_{1,1}=1$) and type-$2$ users are mixed ($\pi_{2,1}\in(0,1)$). For the benevolent provider, we write $\pi^{s,(t_1,t_2)}$ for the NE of
type $(t_1,t_2)$ that maximizes the social welfare, $\mathbf{P}^{s,(t_1,t_2)}$ for the pricing policy that induces $\pi^{s,(t_1,t_2)}$, and
$S(\theta,\alpha,\mathbf{P}^{s,(t_1,t_2)},\pi^{s,(t_1,t_2)})$ ($R(\theta,\alpha,\mathbf{P}^{s,(t_1,t_2)},\pi^{s,(t_1,t_2)})$) for the corresponding
social welfare (revenue). We use the superscript `r', instead of `s', for the counterparts in the case of the selfish provider, e.g., we write
$\pi^{r,(t_1,t_2)}$ for the NE of type $(t_1,t_2)$ that maximizes the revenue.

For the benevolent provider, the type of the optimal NE is determined by
\begin{eqnarray}
(t_1^s,t_2^s) = \arg \max_{(t_1,t_2)} S(\theta,\alpha,\mathbf{P}^{s,(t_1,t_2)},\pi^{s,(t_1,t_2)}).
\end{eqnarray}
Similarly, the type of the optimal NE for the selfish provider is determined by
\begin{eqnarray}
(t_1^r,t_2^r) = \arg \max_{(t_1,t_2)} R(\theta,\alpha,\mathbf{P}^{r,(t_1,t_2)},\pi^{r,(t_1,t_2)}).
\end{eqnarray}
Hence, the provider needs to determine $S(\theta,\alpha,\mathbf{P}^{s,(t_1,t_2)},\pi^{s,(t_1,t_2)})$ or
$R(\theta,\alpha,\mathbf{P}^{r,(t_1,t_2)},\pi^{r,(t_1,t_2)})$ for each type $(t_1,t_2)$. In the following, we will show how the benevolent or selfish
provider determines $S(\theta,\alpha,\mathbf{P}^{s,(t_1,t_2)},\pi^{s,(t_1,t_2)})$ or $R(\theta,\alpha,\mathbf{P}^{r,(t_1,t_2)},\pi^{r,(t_1,t_2)})$
for each type of NE. For convenience, we define $x_k(\pi_{k,1})=\pi_{k,1}\cdot\frac{\lambda_1}{\lambda_1+\mu_1}\frac{p}{1-p}+1$ for $k=1,2$, where
$p$ is the transmission probability in the CSMA protocol.

Before we discuss how to solve the design problem, we derive the analytical expressions of the expected utility of the users as follows.
\begin{lemma}\label{Lemma:ExpressionCSMA}
Suppose that the service provider uses $\theta=\mathrm{CSMA}$ with transmission probability $p$, has no admission control ($\alpha\equiv1$), and
offers the pricing policy ${\bf P} = \left({\bf p}^0 = \phi, {\bf p}^1 = (p_s, 0)\right)$. When the other users choose actions according to the
action profile $\pi$, the expected utility of use and expected cost of a type-$k$ user, whose action is $\pi_k^\prime$, are
\begin{eqnarray}\label{eqn:ExpressionUtilityOfUse_CSMA}
&&U_k(\theta,\alpha,(\pi;\pi_k^\prime)) = \pi_{k,1}^\prime \cdot \Delta T \cdot \frac{\lambda_k}{\lambda_k+\mu_k} \cdot \\
&&\left[\alpha_k- \frac{\beta_k}{p}\left(x_k(\pi_{k,1})\right)^{N_k-1}\left(x_{-k}(\pi_{-k,1})\right)^{N_{-k}}\right] \nonumber
\end{eqnarray}
and $C_k(\theta,\alpha,\mathbf{P},(\pi;\pi_k^\prime)) = \pi_{k,1}^\prime \cdot p_s$.
\end{lemma}
\begin{IEEEproof}
See \cite[Appendix~A]{Appendix}.
\end{IEEEproof}

\subsubsection{Procedures to solve the design problems under CSMA}
Now we show how the providers solve the design problem. Suppose that the service provider uses $\theta=\mathrm{CSMA}$ with transmission probability
$p$, has no admission control ($\alpha\equiv1$), and offers the pricing policy ${\bf P} = \left(\phi, (p_s, 0)\right)$. To maximize the social
welfare, the benevolent provider follows the procedure shown in Table~\ref{Procedure_Benevolent_CSMA}. It compares the social welfare achievable at
six types of NE, and then chooses the optimal NE. At each type of NE, the NE of type $(t_1,t_2)$ that maximizes the social welfare
$\pi^{s,(t_1,t_2)}$, the pricing policy $\mathbf{P}^{s,(t_1,t_2)}$ that induces $\pi^{s,(t_1,t_2)}$, and the corresponding social welfare
$S(\theta,\alpha,\mathbf{P}^{s,(t_1,t_2)},\pi^{s,(t_1,t_2)})$ can be determined either analytically or by solving a convex program as in
Table~\ref{Procedure_Benevolent_CSMA}. To maximize the revenue, the selfish provider compares the revenue achievable at nine types of NE. The
procedure to solve the selfish provider's design problem under CSMA is summarized in Table~\ref{Procedure_Selfish_CSMA}.

\begin{table}
\renewcommand{\arraystretch}{1.3}
\caption{Procedure to solve the benevolent provider's design problem under CSMA.} \label{Procedure_Benevolent_CSMA} \centering
\begin{tabular}{l}
\hline \hline
\textbf{Step 1.} Solve for the case of $(t_1, t_2)=(i,i)$, i.e., both types of users are in: \\
If $\min_{k=1,2} \left\{\Delta T\frac{\lambda_k}{\lambda_k+\mu_k} \left[\alpha_k-\frac{\beta_k}{p}(x_k(1))^{N_k-1}
(x_{-k}(1))^{N_{-k}}\right]\right\}\cdot(N_1+N_2) \ge C_0$, then \\
~~~~$ \pi^{s,(i,i)} = ([0,1],[0,1]),~\mathbf{P}^{s,(i,i)} = \left(\phi,\left(p_s^{s,(i,i)} =\frac{C_0}{N_1+N_2},0\right)\right),$
\\
~~~~$S(\theta,\alpha,\mathbf{P}^{s,(i,i)},\pi^{s,(i,i)}) = \sum_{k=1}^2 \Delta T \frac{N_k \cdot \lambda_k}{\lambda_k+\mu_k}
\left[\alpha_k-\frac{\beta_k}{p}(x_k(1))^{N_k-1} (x_{-k}(1))^{N_{-k}}\right] - C_0; $\\
otherwise, $S(\theta,\alpha,\mathbf{P}^{s,(i,i)},\pi^{s,(i,i)})=-\infty$.
\\
\hline
\textbf{Step 2-3.} Solve for the case of $t_k=i, t_{-k}=o$, i.e., type-$k$ users are in, the other type of users are out ($k=1~\mathrm{or}~2$)~: \\
If $\Delta T\frac{\lambda_k}{\lambda_k+\mu_k} \left[\alpha_k-\frac{\beta_k}{p}(x_k(1))^{N_k-1}\right] \ge \max\left\{\Delta
T\frac{\lambda_{-k}}{\lambda_{-k}+\mu_{-k}} \left[\alpha_{-k}-\frac{\beta_{-k}}{p} (x_{k}(1))^{N_{k}}\right],\frac{C_0}{N_k}\right\}$, then \\
~~~~$\pi_k^{s,(t_k=i, t_{-k}=o)} = [0,1],~\pi_{-k}^{s,(t_k=i, t_{-k}=o)} = [1,0],$ \\
~~~~$\mathbf{P}^{s,(t_k=i, t_{-k}=o)} = \left(\phi, \left(p_s^{s,(t_k=i, t_{-k}=o)}=\max\left\{\Delta T\frac{\lambda_{-k}}{\lambda_{-k}+\mu_{-k}}
\left[\alpha_{-k}-\frac{\beta_{-k}}{p}
(x_{k}(1))^{N_{k}}\right],\frac{C_0}{N_k}\right\},0\right)\right),$ \\
~~~~$S(\theta,\alpha,\mathbf{P}^{s,((t_k=i, t_{-k}=o))},\pi^{s,((t_k=i, t_{-k}=o))}) = \left\{\Delta T \frac{\lambda_k}{\lambda_k+\mu_k}
\left[\alpha_k-\frac{\beta_k}{p}(x_k(1))^{N_k-1}\right]-p_s\right\} \cdot N_k;$ \\
otherwise, $S(\theta,\alpha,\mathbf{P}^{s,((t_k=i, t_{-k}=o))},\pi^{s,((t_k=i, t_{-k}=o))})=-\infty$.
\\
\hline \textbf{Step 4-5.} Solve for the case of $t_k=m, t_{-k}=i$, i.e., type-$k$ users are mixed, the other type of users are in
($k=1~\mathrm{or}~2$)~: \\
$\pi_{k}^{s,(t_k=m, t_{-k}=i)} = [1-\pi_{k,1},\pi_{k,1}],~\pi_{-k}^{s,(t_k=m, t_{-k}=i)}=[0,1]$, where $\pi_{k,1}$ is the solution to the following convex program: \\
$\begin{array}{ccl} S(\theta,\alpha,\mathbf{P}^{s,(t_k=m, t_{-k}=i)},\pi^{s,(t_k=m, t_{-k}=i)}) = &
\displaystyle\max_{\pi_{k,1}\in[0,1]} & \Delta T\frac{\lambda_{-k}}{\lambda_{-k}+\mu_{-k}}\left[\alpha_{-k}-\frac{\beta_{-k}}{p}(x_{-k}(1))^{N_{-k}-1}(x_k(\pi_{k,1}))^{N_k}\right]\cdot N_{-k} \\
& & -\Delta T\frac{\lambda_k}{\lambda_k+\mu_k} \left[\alpha_k-\frac{\beta_k}{p}(x_k(\pi_{k,1}))^{N_k-1}(x_{-k}(1))^{N_{-k}}\right]\cdot N_{-k} \\
&s.t. & \Delta T\frac{\lambda_k}{\lambda_k+\mu_k}
\left[\alpha_k-\frac{\beta_k}{p}(x_k(\pi_{k,1}))^{N_k-1}(x_{-k}(1))^{N_{-k}}\right]\cdot(\pi_{k,1}N_k+N_{-k}) \ge C_0.
\end{array}$ \\
If $S(\theta,\alpha,\mathbf{P}^{s,(t_k=m, t_{-k}=i)},\pi^{s,(t_k=m, t_{-k}=i)})>-\infty$, then \\
~~~~$\mathbf{P}^{s,(t_k=m, t_{-k}=i)} = \left(\phi, \left(p_s^{s,(t_k=m, t_{-k}=i)}=\Delta T\frac{\lambda_k}{\lambda_k+\mu_k}
\left[\alpha_k-\frac{\beta_k}{p}(x_k(\pi_{k,1}))^{N_k-1}(x_{-k}(1))^{N_{-k}}\right],0\right)\right).$
\\
\hline \textbf{Step 6.} Solve for the case of $(t_1, t_2)=(o,o)$, i.e., both types of users are out: \\
$\pi^{s,(o,o)} = ([1,0],[1,0]),~\mathbf{P}^{s,(o,o)} = \left(\phi,
\left(p_s^{s,(o,o)}=\infty,0\right)\right),~S(\theta,\alpha,\mathbf{P}^{s,(o,o)},\pi^{s,(o,o)}) = 0.$
\\
\hline \textbf{Step 7.} Compare $S(\theta,\alpha,\mathbf{P}^{s,(t_1,t_2)},\pi^{s,(t_1,t_2)})$ at the above six types of NE and choose the optimal
NE.\\ \hline \hline
\end{tabular}
\end{table}

%\begin{figure*}[!t]
%\normalsize \setcounter{tempeqncnt}{\value{equation}} \setcounter{equation}{\value{equation}}
%\begin{align}
%\label{eqn:OptimizationProblem_CSMA_Selfish_BothMixed}
%\begin{array}{l}
%R(\theta,\alpha,\mathbf{P}^{r,(m,m)},\pi^{r,(m,m)}) =
%\displaystyle\max_{p_s} p_s\cdot(\pi_{1,1}N_1+\pi_{2,1}N_2) \\
%s.t.~~0\le p_s \le \min_{k=1,2} \left\{\Delta T \frac{\lambda_k}{\lambda_k+\mu_k}\left(\alpha_k-\beta_k/p\right)\right\}, \\
%~~~~~~\pi_{k,1} = \left(\frac{\left[\left(\alpha_{-k}-\frac{p_s}{\Delta T}\frac{\lambda_{-k}+\mu_{-k}}{\lambda_{-k}}\right)/\left(\frac{\beta_{-k}}{p}\right)\right]^{\frac{N_{-k}}{N_1+N_2-1}}}{\left[\left(\alpha_{k}-\frac{p_s}{\Delta T}\frac{\lambda_{k}+\mu_{k}}{\lambda_{k}}\right)/\left(\frac{\beta_{k}}{p}\right)\right]^{\frac{N_{-k}-1}{N_1+N_2-1}}}-1\right)\cdot\frac{\lambda_k+\mu_k}{\lambda_k}\cdot\frac{1-p}{p},~k=1,2, \\
%~~~~~~0\le\pi_{k,1}\le1,~k=1,2, \\
%~~~~~~p_s\cdot(\pi_{1,1}N_1+\pi_{2,1}N_2) \ge C_0.
%\end{array}
%\end{align}
%\hrulefill \vspace*{4pt} \setcounter{equation}{\value{tempeqncnt}+1}
%\end{figure*}

\begin{remark}
To maximize the social welfare, the benevolent provider needs to consider only six types of NE. In contrast, to maximize the revenue, the selfish
provider needs to consider all nine types of NE. This is because for the three types of NE in which no user is in, the social welfare is zero (no
better than not operating the network), but the revenue is positive and is possibly larger than that at the other types of NE. In addition, at the NE
of the type $(m,m)$, which can be neglected by the benevolent provider, the selfish provider has to solve the complicated nonconvex optimization
problem in step~8 of Table~\ref{Procedure_Selfish_CSMA}. To solve this problem, the selfish provider has to exhaustively search for the optimal
subscription fee $p_s^{(m,m)}$. In sum, the computational complexity for the selfish provider to maximize the revenue is higher than that for the
benevolent provider to maximize the social welfare.
\end{remark}

\subsubsection{Comparison between the benevolent and selfish providers}
\begin{figure*}
\centering{\subfloat[]{\includegraphics[width=3.2in]{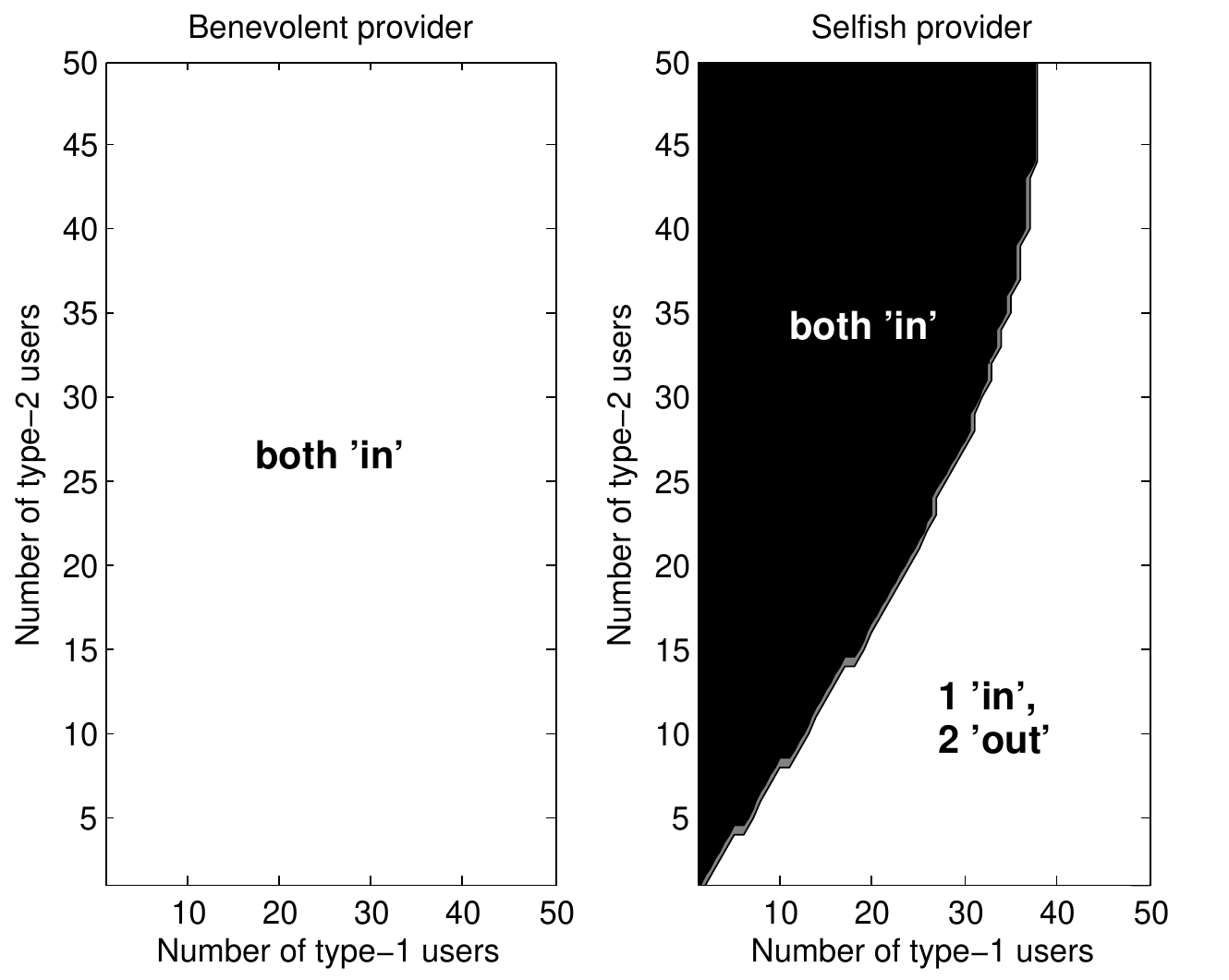}\label{PhaseDiagram_CSMA_MixedStrategy_UserNumber_LL}}
\subfloat[]{\includegraphics[width =3.2in]{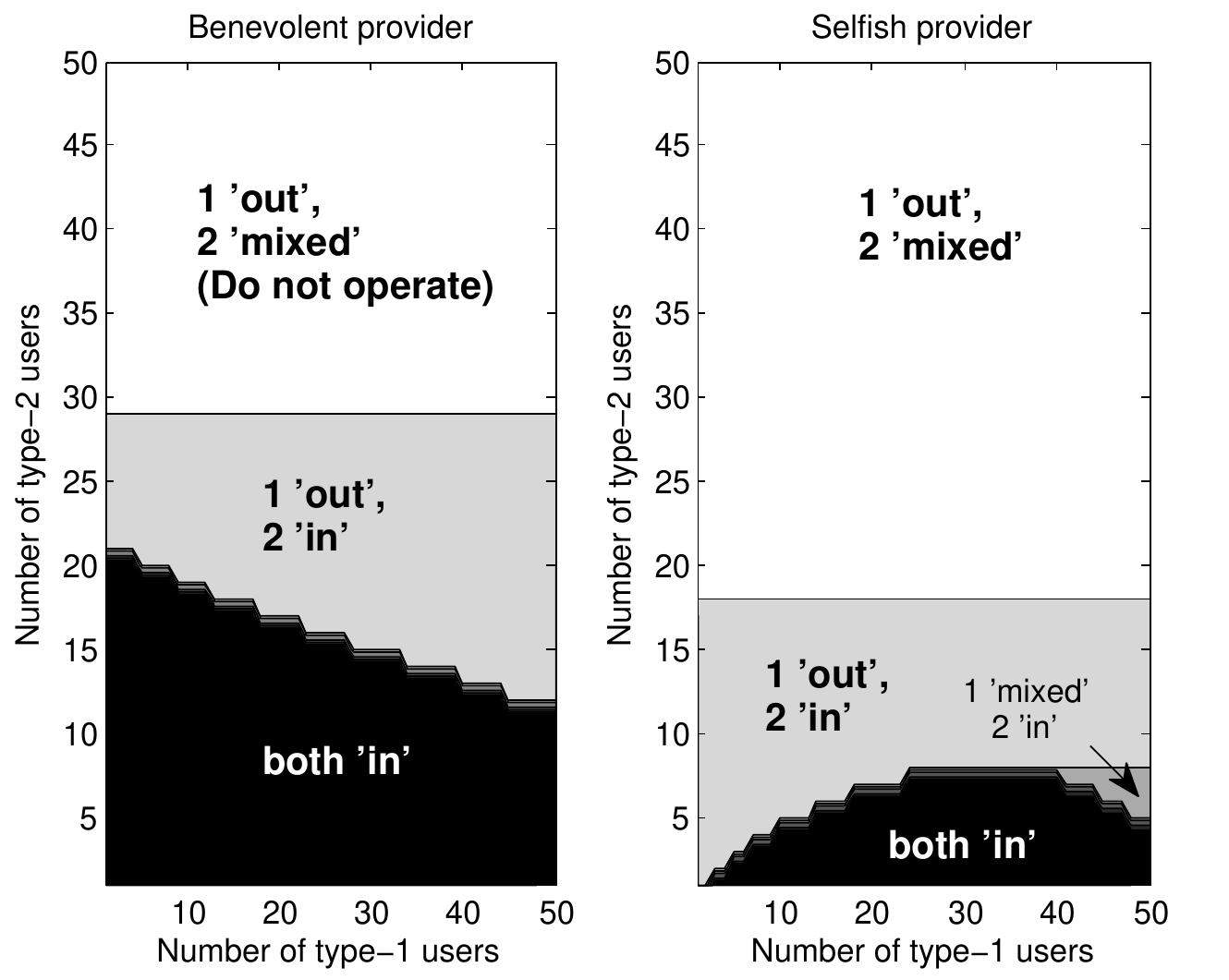}\label{PhaseDiagram_CSMA_MixedStrategy_UserNumber_LH}} \\
\subfloat[]{\includegraphics[width =3.2in]{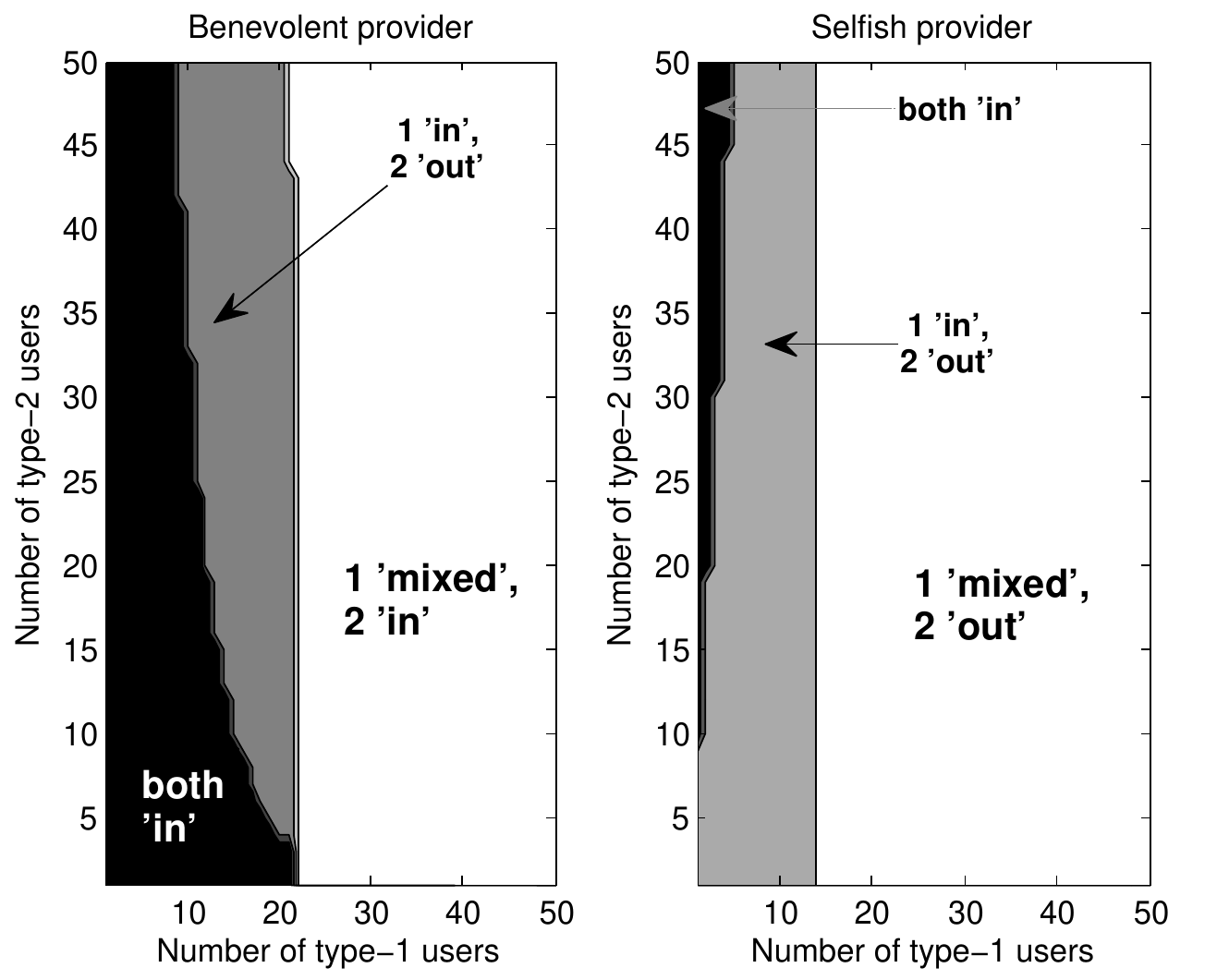}\label{PhaseDiagram_CSMA_MixedStrategy_UserNumber_HL}}
\subfloat[]{\includegraphics[width =3.2in]{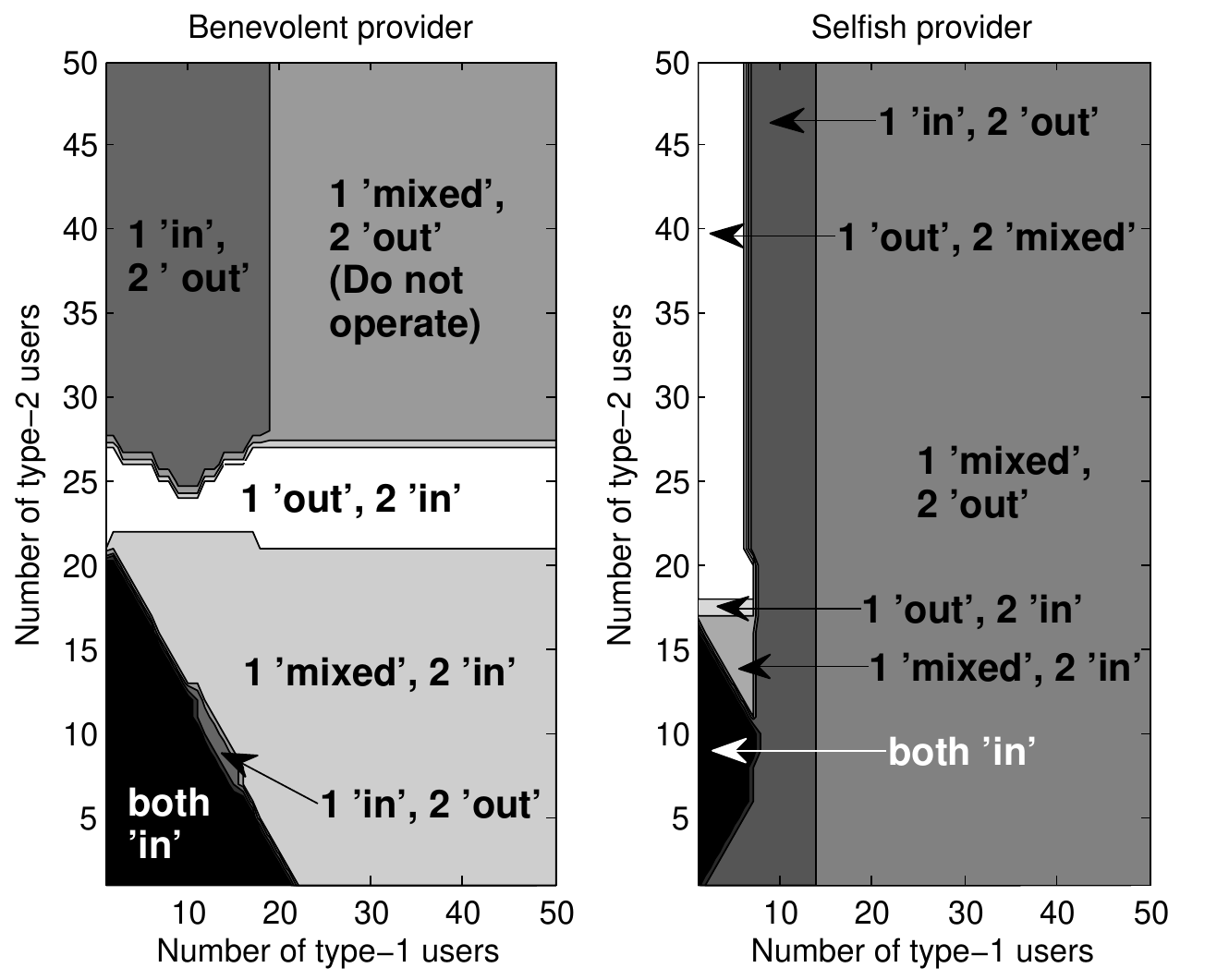}\label{PhaseDiagram_CSMA_MixedStrategy_UserNumber_HH}}}
\caption{Phase diagrams of the optimal types of NE under CSMA. (a): low-demand $\lambda_1/\mu_1=0.1$ video users (type-1) and low-demand
$\lambda_2/\mu_2=0.1$ email users (type-2); (b): low-demand $\lambda_1/\mu_1=0.1$ video users (type-1) and high-demand $\lambda_2/\mu_2=1$ email
users (type-2); (c): high-demand $\lambda_1/\mu_1=1$ video users (type-1) and low-demand $\lambda_2/\mu_2=0.1$ email users (type-2); (d): high-demand
$\lambda_1/\mu_1=1$ video users (type-1) and high-demand $\lambda_2/\mu_2=1$ email users (type-2).} \label{CSMA_PhaseDiagram_FixLambda}
\end{figure*}

As seems obvious, the benevolent provider charges as little as possible, subject to revenue being at least as great as cost; the selfish provider
charges as much as possible, subject to the cost to each user being no greater than utility. Due to the differences in the providers' objectives and
charging schemes, there are ranges of the user number and demand parameters for which the optimal type of NE when the provider is benevolent and the
optimal type when the provider is selfish are different. As an illustration, we show the optimal types of NE under different user number and demand
parameters when the provider is benevolent and selfish in
Fig.~\ref{PhaseDiagram_CSMA_MixedStrategy_UserNumber_LL}--\ref{PhaseDiagram_CSMA_MixedStrategy_UserNumber_HH}. The parameters in the simulation are
as follows:
\begin{itemize}
\item The provider uses a CSMA protocol with a constant backoff window that is equivalent to a slotted ALOHA protocol with transmission probability $p=2/17$.
\item The bandwidth is normalized to $1$.
\item $\alpha_1=10,\beta_1=0.3$ for video users and $\alpha_2=5,\beta_2=0.1$ for email users (same as in Fig.~\ref{ExampleUtility}).
\item The billing period is normalized to $\Delta T=1$.
\item The cost of the service provider is $C_0=0$.
\end{itemize}

\begin{table}
\renewcommand{\arraystretch}{1.3}
\caption{Procedure to solve the selfish provider's design problem under CSMA.} \label{Procedure_Selfish_CSMA} \centering
\begin{tabular}{l}
\hline \hline
\textbf{Step 1.} Solve for the case of $(t_1, t_2)=(i,i)$, i.e., both types of users are in: \\
Define $p_s^{r,(i,i)}\triangleq\min_{k=1,2} \left\{\Delta T\frac{\lambda_k}{\lambda_k+\mu_k} \left[\alpha_k-\frac{\beta_k}{p}(x_k(1))^{N_k-1}
(x_{-k}(1))^{N_{-k}}\right]\right\}$. \\
$\pi^{r,(i,i)} = ([0,1],[0,1]),~\mathbf{P}^{r,(i,i)} = \left(\phi, \left(p_s^{(i,i)},0\right)\right),$
$R(\theta,\alpha,\mathbf{P}^{r,(i,i)},\pi^{r,(i,i)}) = \left\{\begin{array}{ll} p_s^{r,(i,i)}\cdot(N_1+N_2) & \mathrm{if}~p_s^{r,(i,i)}\ge \frac{C_0}{N_1+N_2}\\
-\infty & \mathrm{otherwise}
\end{array}\right..$ \\
\hline
\textbf{Step 2-3.} Solve for the case of $t_k=i, t_{-k}=o$, i.e., type-$k$ users are in, the other type of users are out ($k=1~\mathrm{or}~2$)~: \\
If $\Delta T\frac{\lambda_k}{\lambda_k+\mu_k} \left[\alpha_k-\frac{\beta_k}{p}(x_k(1))^{N_k-1}\right] \ge \max\left\{\Delta
T\frac{\lambda_{-k}}{\lambda_{-k}+\mu_{-k}} \left[\alpha_{-k}-\frac{\beta_{-k}}{p} (x_{k}(1))^{N_{k}}\right],\frac{C_0}{N_k}\right\}$, then \\
~~~~$\pi^{r,(t_k=i, t_{-k}=o)} = ([0,1],[1,0]),~p_s^{r,(t_k=i, t_{-k}=o)}=\Delta T\frac{\lambda_k}{\lambda_k+\mu_k}
\left[\alpha_k-\frac{\beta_k}{p}(x_k(1))^{N_k-1}\right]$, \\
~~~~$R(\theta,\alpha,\mathbf{P}^{r,(t_k=i, t_{-k}=o)},\pi^{r,(t_k=i, t_{-k}=o)}) = \left\{\begin{array}{ll} p_s^{r,(t_k=i, t_{-k}=o)}\cdot N_k & \mathrm{if}~p_s^{r,(t_k=i, t_{-k}=o)}\cdot N_k\ge C_0\\
-\infty & \mathrm{otherwise}
\end{array}\right..$ \\
otherwise, $R(\theta,\alpha,\mathbf{P}^{r,(t_k=i, t_{-k}=o)},\pi^{r,(t_k=i, t_{-k}=o)})=-\infty$. \\
\hline
\textbf{Step 4-5.} Solve for the case of $t_k=m, t_{-k}=i$, i.e., type-$k$ users are mixed, the other type of users are in ($k=1~\mathrm{or}~2$)~: \\
$\pi_{k}^{r,(t_k=m, t_{-k}=i)} = [1-\pi_{k,1},\pi_{k,1}],~\pi_{-k}^{r,(t_k=m, t_{-k}=i)}=[0,1]$, where $\pi_{k,1}$ is the solution to the following convex program: \\
$\begin{array}{cl}
 & R(\theta,\alpha,\mathbf{P}^{r,(t_k=m, t_{-k}=i)},\pi^{r,(t_k=m, t_{-k}=i)}) = \\
\displaystyle\max_{\pi_{k,1}\in[0,1]} & \Delta T\frac{\lambda_k}{\lambda_k+\mu_k}
\left[\alpha_k-\frac{\beta_k}{p}(x_k(\pi_{k,1}))^{N_k-1}(x_{-k}(1))^{N_{-k}}\right]\cdot(\pi_{k,1}N_k+N_{-k}) \\
s.t. & \frac{\lambda_{-k}}{\lambda_{-k}+\mu_{-k}}\left[\alpha_{-k}-\frac{\beta_{-k}}{p}(x_{-k}(1))^{N_{-k}-1}(x_k(\pi_{k,1}))^{N_k}\right] \ge \frac{\lambda_k}{\lambda_k+\mu_k} \left[\alpha_k-\frac{\beta_k}{p}(x_k(\pi_{k,1}))^{N_k-1}(x_{-k}(1))^{N_{-k}}\right], \\
     & \Delta T\frac{\lambda_k}{\lambda_k+\mu_k}
\left[\alpha_k-\frac{\beta_k}{p}(x_k(\pi_{k,1}))^{N_k-1}(x_{-k}(1))^{N_{-k}}\right]\cdot(\pi_{k,1}N_k+N_{-k}) \ge C_0.
\end{array}$ \\
If $R(\theta,\alpha,\mathbf{P}^{r,(t_k=m, t_{-k}=i)},\pi^{r,(t_k=m, t_{-k}=i)})\ge -\infty$, $p_s^{r,(t_k=m, t_{-k}=i)} = \Delta
T\frac{\lambda_k}{\lambda_k+\mu_k}
\left[\alpha_k-\frac{\beta_k}{p}(x_k(\pi_{k,1}))^{N_k-1}(x_{-k}(1))^{N_{-k}}\right]$. \\
\hline
\textbf{Step 6-7.} Solve for the case of $t_k=m, t_{-k}=o$, i.e., type-$k$ users are mixed, the other type of users are out ($k=1~\mathrm{or}~2$)~: \\
$\pi_{k}^{r,(t_k=m, t_{-k}=o)} = [1-\pi_{k,1},\pi_{k,1}],~\pi_{-k}^{r,(t_k=m, t_{-k}=o)}=[1,0]$, where $\pi_{k,1}$ is the solution to the following convex program: \\
$\begin{array}{l} R(\theta,\alpha,\mathbf{P}^{r,(t_k=m, t_{-k}=o)},\pi^{r,(t_k=m, t_{-k}=o)}) = \displaystyle\max_{\pi_{k,1}\in[0,1]} \Delta
T\frac{\lambda_k}{\lambda_k+\mu_k}
\left[\alpha_k-\frac{\beta_k}{p}(x_k(\pi_{k,1}))^{N_k-1}\right]\cdot\pi_{k,1}N_k \\
s.t.~~\Delta T\frac{\lambda_k}{\lambda_k+\mu_k} \left[\alpha_k-\frac{\beta_k}{p}(x_k(\pi_{k,1}))^{N_k-1}\right]\cdot\pi_{k,1}N_k \ge C_0, \\
~~~~~~\frac{\lambda_k}{\lambda_k+\mu_k} \left[\alpha_k-\frac{\beta_k}{p}(x_k(\pi_{k,1}))^{N_k-1}\right] \ge
\frac{\lambda_{-k}}{\lambda_{-k}+\mu_{-k}}\left[\alpha_{-k}-\frac{\beta_{-k}}{p}(x_k(\pi_{k,1}))^{N_k}\right].
\end{array}$ \\
If $R(\theta,\alpha,\mathbf{P}^{r,(t_k=m, t_{-k}=o)},\pi^{r,(t_k=m, t_{-k}=o)})\ge -\infty$, $p_s^{r,(t_k=m, t_{-k}=o)} = \Delta
T\frac{\lambda_k}{\lambda_k+\mu_k}
\left[\alpha_k-\frac{\beta_k}{p}(x_k(\pi_{k,1}))^{N_k-1}\right]$. \\
\hline
\textbf{Step 8.} Solve for the case of $(t_1, t_2)=(m,m)$, i.e., both types of users are mixed:\\
$p_s^{(m,m)}$ is the solution to the following optimization problem: \\
$\begin{array}{ccl} R(\theta,\alpha,\mathbf{P}^{r,(m,m)},\pi^{r,(m,m)}) = &
\displaystyle\max_{p_s} & p_s\cdot(\pi_{1,1}N_1+\pi_{2,1}N_2) \\
&s.t.& 0\le p_s \le \min_{k=1,2} \left\{\Delta T \frac{\lambda_k}{\lambda_k+\mu_k}\left(\alpha_k-\beta_k/p\right)\right\}, \\
&    & \pi_{k,1} = \left(\frac{\left[\left(\alpha_{-k}-\frac{p_s}{\Delta T}\frac{\lambda_{-k}+\mu_{-k}}{\lambda_{-k}}\right)/\left(\frac{\beta_{-k}}{p}\right)\right]^{\frac{N_{-k}}{N_1+N_2-1}}}{\left[\left(\alpha_{k}-\frac{p_s}{\Delta T}\frac{\lambda_{k}+\mu_{k}}{\lambda_{k}}\right)/\left(\frac{\beta_{k}}{p}\right)\right]^{\frac{N_{-k}-1}{N_1+N_2-1}}}-1\right)\cdot\frac{\lambda_k+\mu_k}{\lambda_k}\cdot\frac{1-p}{p},~k=1,2, \\
&    & 0\le\pi_{k,1}\le1,~k=1,2, \\
&    & p_s\cdot(\pi_{1,1}N_1+\pi_{2,1}N_2) \ge C_0.
\end{array}$\\
If $R(\theta,\alpha,\mathbf{P}^{r,(m,m)},\pi^{r,(m,m)})\ge -\infty$, $\pi_{k,1}$ and $\pi_{-k,1}$ are
calculated by the equality constraints in the above optimization problem. \\
\hline
\textbf{Step 9.} Solve for the case of $(t_1, t_2)=(o,o)$, i.e., both types of users are out: \\
$\pi^{r,(o,o)} = ([1,0],[1,0]),~\mathbf{P}^{r,(o,o)} = \left(\phi,
\left(p_s=\infty,0\right)\right),~R(\theta,\alpha,\mathbf{P}^{r,(o,o)},\pi^{r,(o,o)}) = 0.$ \\
\hline \textbf{Step 10.} Compare $R(\theta,\alpha,\mathbf{P}^{r,(t_1,t_2)},\pi^{r,(t_1,t_2)})$ at the above nine types of NE and choose the optimal NE.\\
\hline\hline
\end{tabular}
\end{table}

%\begin{figure}
%\centering
%\includegraphics[width =3.2in]{PhaseDiagram_CSMA_MixedStrategy_UserNumber_LL}
%\caption{Phase diagrams of the optimal types of NE with low-demand $\lambda_1/\mu_1=0.1$ video users (type-1) and low-demand $\lambda_2/\mu_2=0.1$
%email users (type-2) under CSMA.} \label{PhaseDiagram_CSMA_MixedStrategy_UserNumber_LL}
%\end{figure}

Fig.~\ref{PhaseDiagram_CSMA_MixedStrategy_UserNumber_LL} shows the case of low-demand video users and low-demand email users. When both types of
users have low demands, there is a high probability that the number of online users in the system is small. In other words, the congestion level is
low in most of the times. Hence, the utility of use of one user will not decrease significantly with the addition of the other type of users. For
this reason, the benevolent provider prefers the NE in which both types of users are in. This can be done by setting a very low (zero in this case)
subscription such that both types can be in with positive net utilities. For the selfish provider, its revenue at the NE in which both types are in
depends on the smaller one of the utilities of use of the two types of users, namely the utility of use of an email user in this case, and the total
number of users. Since the congestion level is always low, the utility of use of the video users is much larger than that of the email users in most
of the times. Hence, in most cases, the selfish provider prefers the NE in which only video users are in, because it can set a subscription fee
almost as high as the utility of use of a video user. However, it will set a smaller subscription fee to let both types in, when the number of email
users is much larger than that of the video users.

%\begin{figure}
%\centering
%\includegraphics[width =3.2in]{PhaseDiagram_CSMA_MixedStrategy_UserNumber_LH}
%\caption{Phase diagrams of the optimal types of NE with low-demand $\lambda_1/\mu_1=0.1$ video users (type-1) and high-demand $\lambda_2/\mu_2=1$
%email users (type-2) under CSMA.} \label{PhaseDiagram_CSMA_MixedStrategy_UserNumber_LH}
%\end{figure}

Fig.~\ref{PhaseDiagram_CSMA_MixedStrategy_UserNumber_LH} shows the case of low-demand video users and high-demand email users. Although the demand of
email users is high, the congestion level is still always low if the number of email users is small. Hence, the benevolent provider sets a low
subscription fee to let both types in when the number of email users is small. When the number of email users is higher, the utility of use of video
users decreases to below zero, because they need high throughput to get a positive utility of use. Hence, video users choose to be out when the
number of email users is high (around 25). When the number of data users keeps growing, they has to be mixed to reduce the congestion level. Since
the video users are out and the data users are mixed, the social welfare is zero, the same as in the case when the provider does not setup the
network. Since the utility of use of video users is low even when the number of data users is small, the selfish provider selects the NE in which
both are in only when the video users outnumber the email users. When the number of data users is large, the utility of use of video users is
negative, and the selfish provider will choose the NE in which video users are out.

Fig.~\ref{PhaseDiagram_CSMA_MixedStrategy_UserNumber_HL} shows the case of high-demand video users and low-demand email users. This case is similar
to the previous one, except that the high-demand users are the video users instead of the email users. Both providers choose the NE in which both
types are in when there are a small number of high-demand video users. When the number of video users grows, they choose the NE in which email users
are out. The difference between these two cases happens when the number of high-demand users is very high and they choose to be mixed. In this case,
the low-demand email users can still be in the network with positive utilities, because they do not require high throughput to have positive
utilities. On the contrary, in the previous case, the low-demand video users can not get positive utilities in the network even when the high-demand
email users choose to be mixed. It is worth noticing that in the case of high-demand video users and low-demand email users, in order to maximize the
social welfare, the benevolent provider may charge a positive subscription fee even when the cost is zero. This happens when the number of video
users is around 20 and that of email users is around 5. Intuitively, the benevolent provider should not charge to make profit. However, it has to
make profit to maximize the social welfare in certain circumstances.

Fig.~\ref{PhaseDiagram_CSMA_MixedStrategy_UserNumber_HH} shows the case of high-demand video users and high-demand email users. In this case, the
congestion level is high as long as there is at least one type of users with a large user number. Hence, both providers tend to choose the NE in
which only one type is in, except when both types have small user numbers. Roughly speaking, they prefer the NE in which the type of users with a
smaller user number are in, such that the utility of use of each online user is high.

\begin{remark}
As we have discussed in Remark~1, the selfish provider needs to undertake high computational complexity to determine the highest revenue achievable
at the NE in which both types are mixed. If the computational complexity is beyond its limit, the selfish provider can neglect the NE of the type
$(m,m)$ to get a suboptimal revenue. In a wide range of user number and demand parameters shown in
Fig.~\ref{PhaseDiagram_CSMA_MixedStrategy_UserNumber_LL}--\ref{PhaseDiagram_CSMA_MixedStrategy_UserNumber_HH}, it will not cause any loss in the
revenue by neglecting the NE of the type $(m,m)$. Moreover, the providers only need to solve the design problems once before the network is setup
\end{remark}

\subsection{TDMA}
The provider using TDMA offers the dummy pricing plan ${\bf p}^0 = \phi$ and non-dummy pricing plans that charge subscription fees and per-bit rates
${\bf p} = (p_s, q)$. Again, we use backward induction to analyze the design problem of the provider. The benevolent (selfish) provider compares the
social welfare (revenue) at all the possible equilibria of the plan selection game and chooses the NE with the highest social welfare (revenue). We
first prove that it is sufficient to consider the pricing policies that consist of a dummy pricing plan and one non-dummy pricing plan.
\begin{proposition}\label{proposition:EquivalentPricingPolicy}
Suppose that the benevolent service provider uses $\theta=\mathrm{TDMA}$, has no admission control ($\alpha\equiv1$), and offers the pricing policy
${\bf P}^\prime$. For any NE $\pi$ of the plan selection game $\mathcal{G}_{\mathbf{P}^\prime}$, we can find a pricing policy $\mathbf{P}=(\phi,
(p_s,q))$, such that $\pi$ is a NE of the plan selection game $\mathcal{G}_{\mathbf{P}}$ and that $C_k(\theta,\alpha,\mathbf{P},\pi) =
C_k(\theta,\alpha,\mathbf{P}^\prime,\pi)$ for $k=1,2$.
\end{proposition}
\begin{IEEEproof}
See \cite[Appendix~C]{Appendix}.
\end{IEEEproof}

\begin{table}
\renewcommand{\arraystretch}{1.3}
\caption{Procedure to solve the benevolent provider's design problem under TDMA.} \label{Procedure_Benevolent_TDMA} \centering
\begin{tabular}{l}
\hline \hline
\textbf{Step 1.} Solve for the case of $(t_1, t_2)=(i,i)$, i.e., both types of users \\
~~~~~~~~~~are in: \\
Define $\pi^{s,(i,i)} = ([0,1],[0,1])$, $k=\displaystyle\min_{m=1,2} U_m(\theta,\alpha,\pi^{s,(i,i)})$, and \\
$\rho=\min\left\{\frac{U_{-k}(\theta,\alpha,\pi^{s,(i,i)})}{U_{k}(\theta,\alpha,\pi^{s,(i,i)})},\max\left\{1,\frac{B_{-k}(\theta,\alpha,\pi^{s,(i,i)})}{B_{k}(\theta,\alpha,\pi^{s,(i,i)})}\right\}\right\}$. \\
If $U_{k}(\theta,\alpha,\pi^{s,(i,i)})\cdot (N_{k} + \rho \cdot N_{-k}) \ge C_0$, then \\
~~~~$p_s^{s,(i,i)} = \frac{C_0}{\rho \cdot
N_{-k}+N_k}\left(1-\frac{\rho-1}{\frac{B_{-k}(\theta,\alpha,\pi^{s,(i,i)})}{B_{k}(\theta,\alpha,\pi^{s,(i,i)})}-1}\right),$\\
~~~~$q^{s,(i,i)} = \frac{C_0}{\rho \cdot N_{-k}+N_k}\cdot\frac{\rho-1}{B_{-k}(\theta,\alpha,\pi^{s,(i,i)})-B_{k}(\theta,\alpha,\pi^{s,(i,i)})},$ \\
~~~~$S(\theta,\alpha,\mathbf{P}^{s,(i,i)},\pi^{s,(i,i)}) = \sum_{k=1}^2 U_k(\theta,\alpha,\pi^{s,(i,i)})\cdot N_k - C_0;$ \\
otherwise, $S(\theta,\alpha,\mathbf{P}^{s,(i,i)},\pi^{s,(i,i)})=-\infty$. \\
\hline
\textbf{Step 2-3.} Solve for the case of $t_k=i, t_{-k}=o$, i.e., type-$k$ users are in, \\
~~~~~~~~~~~~the other type of users are out ($k=1~\mathrm{or}~2$)~: \\
Define $\pi_k^{s,(t_k=i, t_{-k}=o)} = [0,1]$, $\pi_{-k}^{s,(t_k=i, t_{-k}=o)}=[1,0]$, and $\rho=$\\
$\max\!\!\left\{\frac{C_0/N_k}{U_{-k}(\theta,\alpha,\pi^{s,(t_k=i, t_{-k}=o)})},\min\!\!\left\{1,\!\frac{B_{k}(\theta,\alpha,\pi^{s,(t_k=i, t_{-k}=o)})}{B_{-k}(\theta,\alpha,\pi^{s,(t_k=i, t_{-k}=o)})}\right\}\!\!\right\}$ \\
If $U_{k}(\theta,\alpha,\pi^{s,(t_k=i, t_{-k}=o)}) \ge \rho \cdot U_{-k}(\theta,\alpha,\pi^{s,(t_k=i, t_{-k}=o)})$, then \\
~~~~$\begin{array}{ccl} p_s^{s,(t_k=i, t_{-k}=o)} &=& U_{-k}(\theta,\alpha,\pi^{s,(t_k=i, t_{-k}=o)}) \\ &\cdot& \left(1-\frac{1-\min\{1,\rho\}}{1-\frac{B_{k}(\theta,\alpha,\pi^{s,(t_k=i, t_{-k}=o)})}{B_{-k}(\theta,\alpha,\pi^{s,(t_k=i, t_{-k}=o)})}}\right)\end{array},$ \\
\\~~~~$\begin{array}{cl}&q^{s,(t_k=i, t_{-k}=o)} \\ = & \!\!\!\!U_{-k}(\theta,\alpha,\pi^{s,(t_k=i, t_{-k}=o)}) \\ \cdot
&\!\!\!\!\frac{1-\min\{1,\rho\}}{B_{-k}(\theta,\alpha,\pi^{s,(t_k=i, t_{-k}=o)})-B_{k}(\theta,\alpha,\pi^{s,(t_k=i, t_{-k}=o)})}\end{array},$ \\
\\~~~~$\begin{array}{cl}&S(\theta,\alpha,\mathbf{P}^{s,(t_k=i, t_{-k}=o)},\pi^{s,(t_k=i, t_{-k}=o)}) \\ = & \!\!\!\!\!\!\left(U_{k}(\theta,\alpha,\pi^{s,(t_k=i,
t_{-k}=o)}) - \rho \cdot
U_{-k}(\theta,\alpha,\pi^{s,(t_k=i, t_{-k}=o)})\right) \\ \cdot & \!\!\!\!N_k \end{array}$ \\
otherwise, $S(\theta,\alpha,\mathbf{P}^{s,(t_k=i, t_{-k}=o)},\pi^{s,(t_k=i, t_{-k}=o)})=-\infty$. \\
\hline
\textbf{Step 4-5.} Solve for the case of $t_k=m, t_{-k}=i$, i.e., type-$k$ users are \\
~~~~~~~~~~~~mixed, the other type of users are in
($k=1~\mathrm{or}~2$)~: \\
Exhaustively search for $\pi_{k,1}$ in $\pi^{s,(t_k=m, t_{-k}=i)}$. \\
\hline \textbf{Step 6.} Solve for the case of $(t_1, t_2)=(o,o)$, i.e., both types of users \\
~~~~~~~~~~are out: \\
$\pi^{s,(o,o)} = ([1,0],[1,0]),~p_s^{s,(o,o)}=\infty,~q^{s,(o,o)}=0,$\\
$S(\theta,\alpha,\mathbf{P}^{s,(o,o)},\pi^{s,(o,o)}) = 0.$
\\
\hline \textbf{Step 7.} Compare $S(\theta,\alpha,\mathbf{P}^{s,(t_1,t_2)},\pi^{s,(t_1,t_2)})$ at the above six types \\
~~~~~~~~~~of NE and choose the optimal NE.\\ \hline \hline
\end{tabular}
\end{table}

Proposition~\ref{proposition:EquivalentPricingPolicy} allows the providers to offer a simple pricing policy $\mathbf{P}=(\phi, (p_s,q))$ that
consists of a dummy pricing plan and a single non-dummy pricing plan, without sacrificing the social welfare or the revenue. It also simplifies our
following analysis. Similar to the case of CSMA, there are nine types of NE in the plan selection game, depending on whether a type of users are in,
out, or mixed. Hence, we can use the same superscript $(t_1,t_2)$ to denote the type of NE.

Before solving the design problems, we derive the analytical expressions of the expected utility of use and the expected cost as follows.
\begin{lemma}\label{Lemma:ExpressionTDMA}
Suppose that the service provider uses $\theta=\mathrm{TDMA}$, has no admission control ($\alpha\equiv1$), and offers the pricing policy ${\bf P} =
\left({\bf p}^0 = \phi, {\bf p}^1 = (p_s, q)\right)$. When the other users choose actions according to the action profile $\pi$, the expected utility
of use and expected cost of a type-$k$ user, whose action is $\pi_k^\prime$, are
\begin{eqnarray}\label{eqn:ExpressionUtilityOfUse_TDMA}
&&U_k(\theta,\alpha,(\pi;\pi_k^\prime)) = \pi_{k,1}^\prime \cdot \Delta T \frac{\lambda_k}{\lambda_k+\mu_k} \cdot\\
&&\left[\alpha_k-\beta_k\left(1+\frac{\lambda_k(N_k-1)}{\lambda_k+\mu_k}\pi_{k,1}+\frac{\lambda_{-k}N_{-k}}{\lambda_{-k}+\mu_{-k}}\pi_{-k,1}\right)\right]
\nonumber
\end{eqnarray}
and
\begin{eqnarray}\label{eqn:ExpressionCost_TDMA}
C_k(\theta,\alpha,\mathbf{P},(\pi;\pi_k^\prime)) = \pi_{k,1}^\prime \cdot \left(p_s + q\cdot \hat{B}_k(\theta,\alpha,\pi)\right),
\end{eqnarray}
where $\hat{B}_k(\theta,\alpha,\pi)$ is the expected data usage of an online type-$k$ user calculated as
\begin{eqnarray}\label{eqn:ExpressionDataRate_TDMA}
\hat{B}_k(\theta,\alpha,\pi) = \nonumber \\
\sum_{n_{k,1}=1}^{N_k} \sum_{n_{-k,1}=0}^{N_{-k}} \binom {N_k-1}{n_{k,1}-1}\cdot\pi_{k,1}^{n_{k,1}-1} (1-\pi_{k,1})^{N_k-n_{k,1}} \nonumber \\
\cdot \binom {N_{-k}}{n_{-k,1}}\cdot\pi_{-k,1}^{n_{-k,1}} (1-\pi_{-k,1})^{N_{-k}-n_{-k,1}} \cdot B_k^1(\theta,\alpha,{\bf n}),
\end{eqnarray}
where $B_k^1(\theta,\alpha,{\bf n})$ is calculated in \eqref{DataRateGivenUserNumber} with $\tau^\theta_{k,1}({\bf x})=\frac{1}{x_{k,1}+x_{-k,1}}$.
\end{lemma}
\begin{IEEEproof}
See \cite[Appendix~D]{Appendix}.
\end{IEEEproof}
%As we can see from the above lemma, the analytical expression for the expected cost is complicated, because of the complicated expression for the
%data rate. Hence, there may be no efficient algorithm in computing the highest social welfare or revenue achievable at mixed NE.

\subsubsection{Procedures to solve the design problems under TDMA}
Suppose that the service provider uses $\theta=\mathrm{TDMA}$, has no admission control ($\alpha\equiv1$), and offers the pricing policy ${\bf P} =
\left(\phi, (p_s, q)\right)$. To maximize the social welfare, the benevolent provider follows the procedure shown in
Table~\ref{Procedure_Benevolent_TDMA}. It compares the social welfare achievable at six types of NE, and then chooses the optimal NE. Likewise, to
maximize the revenue, the selfish provider compares the revenue achievable at nine types of NE. The procedure to solve the selfish provider's design
problem under TDMA is summarized in Table~\ref{Procedure_Selfish_TDMA}.

\begin{table}
\renewcommand{\arraystretch}{1.3}
\caption{Procedure to solve the selfish provider's design problem under TDMA.} \label{Procedure_Selfish_TDMA} \centering
\begin{tabular}{l}
\hline \hline
\textbf{Step 1.} Solve for the case of $(t_1, t_2)=(i,i)$, i.e., both types of users are in: \\
Define $\pi^{r,(i,i)} = ([0,1],[0,1])$, $k=\displaystyle\min_{m=1,2} U_m(\theta,\alpha,\pi^{r,(i,i)})$, and \\
$\rho=\min\left\{\frac{U_{-k}(\theta,\alpha,\pi^{r,(i,i)})}{U_{k}(\theta,\alpha,\pi^{r,(i,i)})},\max\left\{1,\frac{B_{-k}(\theta,\alpha,\pi^{r,(i,i)})}{B_{k}(\theta,\alpha,\pi^{r,(i,i)})}\right\}\right\}$. \\
If $U_{k}(\theta,\alpha,\pi^{r,(i,i)})\cdot (N_{k} + \rho \cdot N_{-k}) \ge C_0$, then \\
~~~~$p_s^{r,(i,i)} = U_{k}(\theta,\alpha,\pi^{r,(i,i)})\left(1-\frac{\rho-1}{\frac{B_{-k}(\theta,\alpha,\pi^{r,(i,i)})}{B_{k}(\theta,\alpha,\pi^{r,(i,i)})}-1}\right),~q^{r,(i,i)} = U_{k}(\theta,\alpha,\pi^{r,(i,i)})\cdot\frac{\rho-1}{B_{-k}(\theta,\alpha,\pi^{r,(i,i)})-B_{k}(\theta,\alpha,\pi^{r,(i,i)})},$ \\
~~~~$R(\theta,\alpha,\mathbf{P}^{r,(i,i)},\pi^{r,(i,i)}) = U_{k}(\theta,\alpha,\pi^{r,(i,i)})\cdot (N_{k} + \rho \cdot N_{-k})$ \\
otherwise, $R(\theta,\alpha,\mathbf{P}^{r,(i,i)},\pi^{r,(i,i)})=-\infty$.\\
\hline
\textbf{Step 2-3.} Solve for the case of $t_k=i, t_{-k}=o$, i.e., type-$k$ users are in, the other type of users are out ($k=1~\mathrm{or}~2$)~: \\
Define $\pi_k^{r,(t_k=i, t_{-k}=o)} = [0,1]$, $\pi_{-k}^{r,(t_k=i, t_{-k}=o)}=[1,0]$, and \\
If $U_{k}(\theta,\alpha,\pi^{r,(t_k=i, t_{-k}=o)}) \ge \rho \cdot U_{-k}(\theta,\alpha,\pi^{r,(t_k=i, t_{-k}=o)})$, then \\
~~~~$p_s^{r,(t_k=i, t_{-k}=o)} = U_{-k}(\theta,\alpha,\pi^{r,(t_k=i, t_{-k}=o)})\left(\rho-\frac{1-\min\{1,\rho\}}{\frac{B_{-k}(\theta,\alpha,\pi^{r,(t_k=i, t_{-k}=o)})}{B_{k}(\theta,\alpha,\pi^{r,(t_k=i, t_{-k}=o)})}-1}\right),$ \\
~~~~$q^{r,(t_k=i, t_{-k}=o)} = U_{-k}(\theta,\alpha,\pi^{r,(t_k=i, t_{-k}=o)})\cdot\frac{1-\min\{1,\rho\}}{B_{-k}(\theta,\alpha,\pi^{r,(t_k=i,
t_{-k}=o)})-B_{k}(\theta,\alpha,\pi^{r,(t_k=i, t_{-k}=o)})},$ \\
~~~~$R(\theta,\alpha,\mathbf{P}^{r,(t_k=i, t_{-k}=o)},\pi^{r,(t_k=i, t_{-k}=o)}) = \rho \cdot U_{-k}(\theta,\alpha,\pi^{r,(t_k=i, t_{-k}=o)})) \cdot
N_k;$ \\
otherwise, $R(\theta,\alpha,\mathbf{P}^{r,(t_k=i, t_{-k}=o)},\pi^{r,(t_k=i, t_{-k}=o)})=-\infty$. \\
\hline
\textbf{Step 4-5.} Solve for the case of $t_k=m, t_{-k}=i$, i.e., type-$k$ users are mixed, the other type of users are in ($k=1~\mathrm{or}~2$)~: \\
Exhaustively search for $\pi_{k,1}$ in $\pi^{r,(t_k=m, t_{-k}=i)}$. \\
\hline
\textbf{Step 6-7.} Solve for the case of $t_k=m, t_{-k}=o$, i.e., type-$k$ users are mixed, the other type of users are out ($k=1~\mathrm{or}~2$)~: \\
Exhaustively search for $\pi_{k,1}$ in $\pi^{r,(t_k=m, t_{-k}=o)}$. \\
\hline
\textbf{Step 8.} Solve for the case of $(t_1, t_2)=(m,m)$, i.e., both types of users are mixed: \\
Exhaustively search for $\pi_{k,1}$ and $\pi_{-k,1}$ in $\pi^{r,(m,m)}$. \\
\hline
\textbf{Step 9.} Solve for the case of $(t_1, t_2)=(o,o)$, i.e., both types of users are out: \\
$\pi^{r,(o,o)} = ([1,0],[1,0]),~p_s^{r,(o,o)}=\infty,~q^{r,(o,o)}=0,~S(\theta,\alpha,\mathbf{P}^{r,(o,o)},\pi^{r,(o,o)}) = 0.$ \\
\hline \textbf{Step 10.} Compare $R(\theta,\alpha,\mathbf{P}^{r,(t_1,t_2)},\pi^{r,(t_1,t_2)})$ at the above nine types of NE and choose the optimal NE.\\
\hline\hline
\end{tabular}
\end{table}

\begin{remark}
Using TDMA enables the provider to charge users for the guaranteed data rates, which results in higher social welfare or revenue. However, as we can
see from Table~\ref{Procedure_Benevolent_TDMA} and Table~\ref{Procedure_Selfish_TDMA}, the provider has to perform exhaustive search for the highest
social welfare or revenue achievable at certain types of NE, due to the complicated expression for the expected data rate
\eqref{eqn:ExpressionDataRate_TDMA}. More specifically, the provider can analytically solve the cases of the four types of NE with no user being
mixed, but needs exhaustive search for the cases of the five types of NE in which at least one type of users are mixed. Hence, to achieve optimal
performance, the provider using TDMA has a higher computational complexity than the one using CSMA. However, the additional complexity is acceptable,
because the provider solves the design problem only once prior to the setup of the network.
\end{remark}

\subsubsection{Comparison between CSMA and TDMA}
We illustrate the solutions to the design problems when the providers use TDMA. Since the providers' objectives do not change in TDMA, the difference
in the behaviors of the benevolent provider and the selfish provider is similar to the case when they use CSMA, which has been discussed in the
previous subsection. Hence, the focus here is the comparison between CSMA and TDMA in terms of the solutions to the design problems and the resulting
social welfare and revenue.

\begin{figure}
\centering
\includegraphics[width =3.5in]{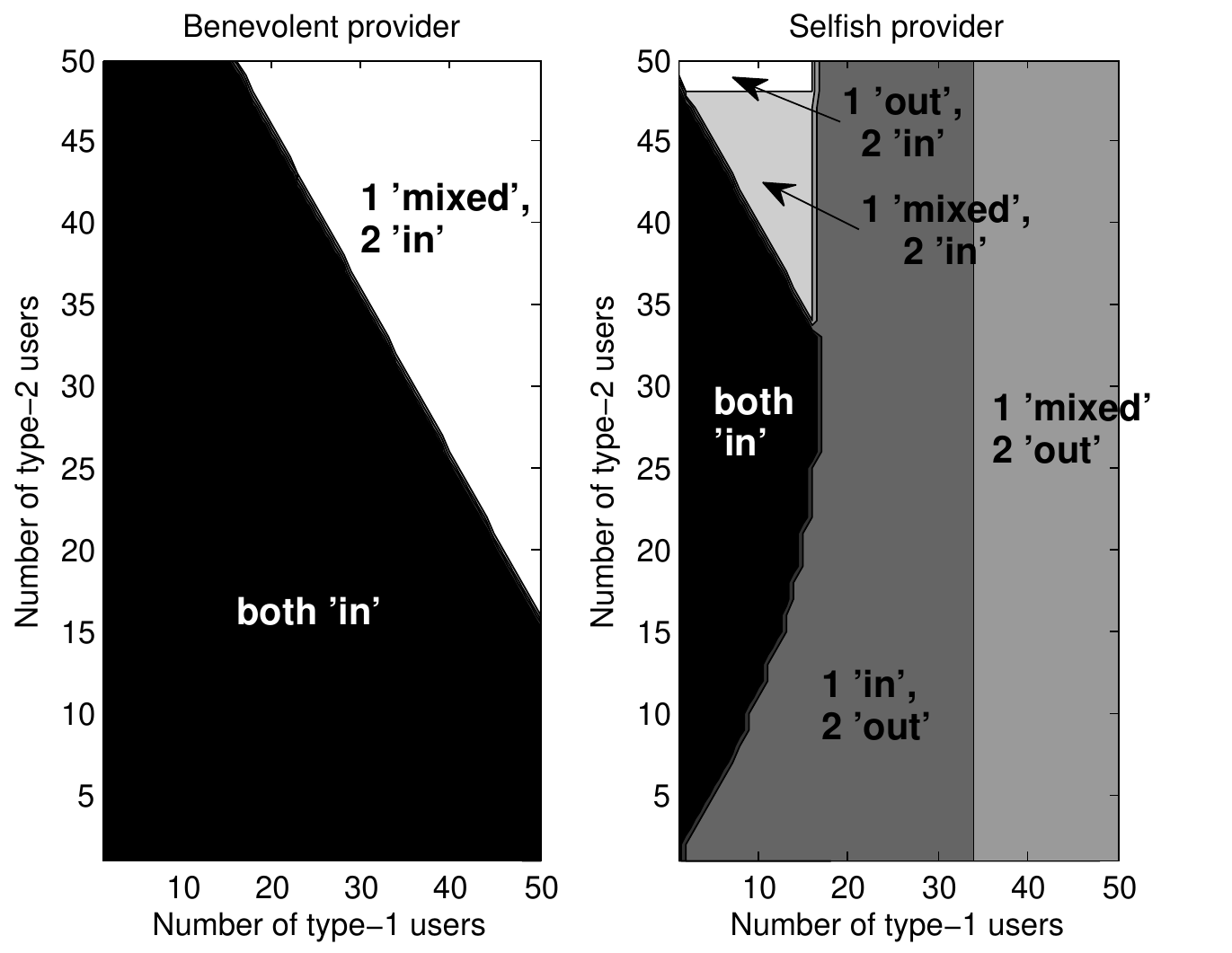}
\caption{Phase diagrams of the optimal types of NE with high-demand $\lambda_1/\mu_1=1$ video users (type-1) and high-demand $\lambda_2/\mu_2=1$
email users (type-2) under TDMA.} \label{PhaseDiagram_TDMA_MixedStrategy_UserNumber_HH}
\end{figure}

First, we show the phase diagram of the optimal types of NE with high-demand video and email users under TDMA in
Fig.~\ref{PhaseDiagram_TDMA_MixedStrategy_UserNumber_HH}. The parameters in the simulation are the same as those in the previous subsection. Compared
to the phase diagram under CSMA in Fig.~\ref{PhaseDiagram_CSMA_MixedStrategy_UserNumber_HH}, we can see that for both the benevolent and selfish
providers, it is more likely that they choose the pricing policies such that both types are in the network, and is less likely that they choose the
pricing policies such that one type are out of the network. This is because under the same number of users, the congestion under TDMA is smaller than
that under CSMA. The difference between the congestion under TDMA and that under CSMA is large especially when the number of users is large, as can
be seen from the expressions of the utility of use in \eqref{eqn:ExpressionUtilityOfUse_CSMA} of Lemma~\ref{Lemma:ExpressionCSMA} and
\eqref{eqn:ExpressionUtilityOfUse_TDMA} of Lemma~\ref{Lemma:ExpressionTDMA}: the utility of use under CSMA decreases exponentially with the number of
users, while the utility of use under TDMA decreases linearly with the number of users. Since the congestion under TDMA is smaller, the providers can
obtain high social welfare and revenue at the NE in which both types are in the network. Note that in
Fig.~\ref{PhaseDiagram_TDMA_MixedStrategy_UserNumber_HH}, we assume that the cost is zero, which is the same as the cost of using CSMA. However, it
is more reasonable to assume that the cost of using TDMA is higher than that of using CSMA. In the following, we compare the social welfare and the
revenue when using TDMA and CSMA under different user number and demand parameters, considering the cost difference of TDMA and CSMA.

\begin{figure}
\centering
\includegraphics[width =3.5in]{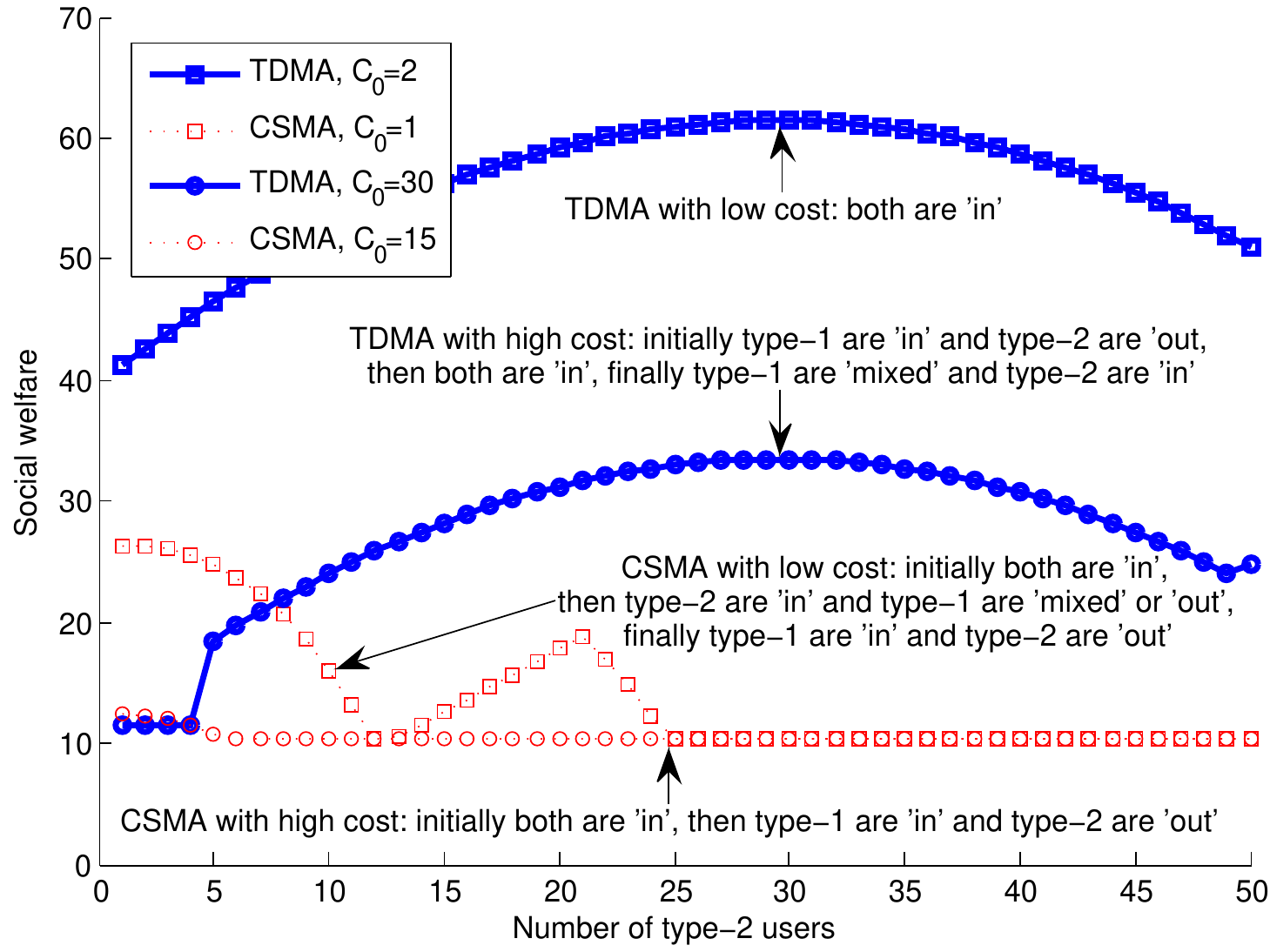}
\caption{Optimal social welfare achieved by the benevolent provider with high-demand $\lambda_1/\mu_1=1$ video users (type-1) and high-demand
$\lambda_2/\mu_2=1$ email users (type-2) under CSMA and TDMA. The number of video users is 10, and the number of email users grows from 1 to 50.}
\label{CSMAvsTDMA_SocialWelfare_DemandHH_10VideoUsers}
\end{figure}

\begin{figure}
\centering
\includegraphics[width =3.5in]{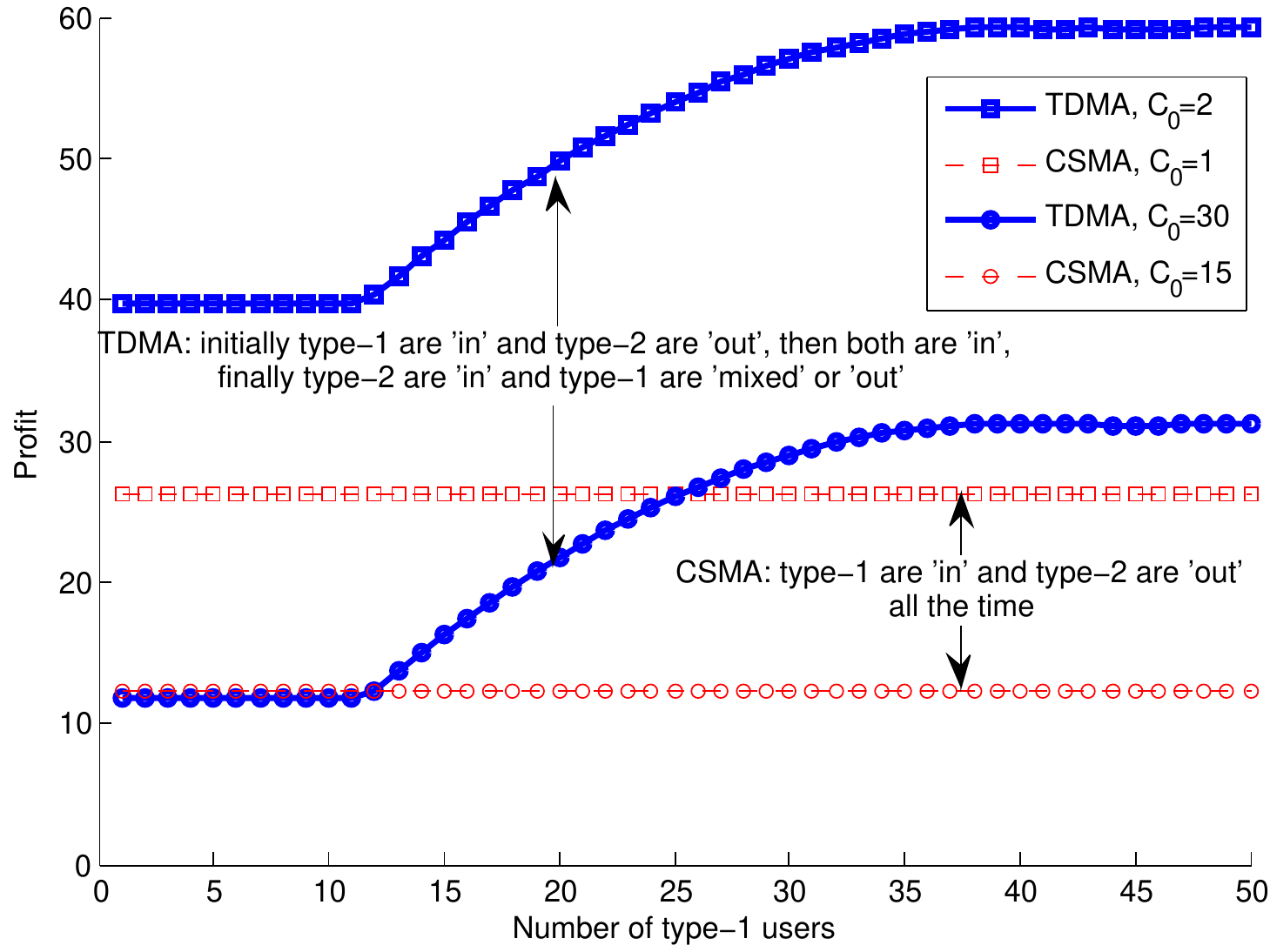}
\caption{Optimal profit (revenue) achieved by the selfish provider with high-demand $\lambda_1/\mu_1=1$ video users (type-1) and high-demand
$\lambda_2/\mu_2=1$ email users (type-2) under CSMA and TDMA. The number of video users is 10, and the number of email users grows from 1 to 50.}
\label{CSMAvsTDMA_Profit_DemandHH_10VideoUsers}
\end{figure}

Fig.~\ref{CSMAvsTDMA_SocialWelfare_DemandHH_10VideoUsers} shows the optimal social welfare achieved by the benevolent provider under CSMA and TDMA.
We can see that when the cost difference is low ($C_0=1$ for CSMA and $C_0=2$ for TDMA), TDMA is always better in the range of email user number
parameters considered. When the cost difference is high ($C_0=15$ for CSMA and $C_0=30$ for TDMA), TDMA is worse initially, when it does not cover
the cost to have both types in the network due to the small number of email users. Similarly, as shown in
Fig.~\ref{CSMAvsTDMA_Profit_DemandHH_10VideoUsers}, the profit achieved by the selfish provider under TDMA is always better when the cost difference
is low, and that achieved under TDMA could be worse when the cost difference is high and the number of email users is small. The providers can also
draw such figures under different user number and demand parameters and under different costs for CSMA and TDMA, in order to predetermine which MAC
protocol to adopt.

\section{Conclusion}
In this paper, we studied the provision of a wireless network by a monopolistic provider who may be either benevolent (seeking to maximize social
welfare) or selfish (seeking to maximize revenue). The paper presented a model for the public wireless network with three interdependent layers,
namely the technology layer, the application layer, and the economic layer. Using the proposed model, we analyzed the influence of technology on the
economic layer, and more importantly, the interaction of technology and economic layers that determines the feasibility and desirability of the
network. We derived the social welfare (the revenue) and the corresponding optimal pricing policy at the optimal operating points of the benevolent
(selfish) service providers for the wireless network under different technologies. By simulation, we characterized different behaviors of a
benevolent provider and a selfish provider at their optimal operating points, and the difference social welfare and revenue resulting from the
different behaviors. Simulation results also demonstrated that differences in MAC technology can have a significant effect on the system performance.
By using TDMA, which enables the providers to charge per-bit rate, both the benevolent provider and the selfish provider can exploit the flexibility
of differentiated pricing plans in order to maximize social welfare and revenue, respectively.

\appendices

\end{document}